\documentclass[10pt,journal]{IEEEtran}
\usepackage{cite} 
\usepackage{url}  
\usepackage{ifthen}  
\usepackage{multicol}   
\usepackage[utf8]{inputenc} 
\usepackage{moreverb}
\usepackage{graphicx}
\usepackage{amsmath,amsfonts,amssymb}
\usepackage{epsfig,subfig,algorithm,algorithmic}
\usepackage{epstopdf}
\usepackage{url}
\usepackage{subfig}
\usepackage{url,amsmath,amssymb}

\begin{document}
%
\title{Aerial-Terrestrial Communications: Terrestrial Cooperation and Energy-Efficient \\Transmissions to Aerial-Base Stations}

\author{Sithamparanathan Kandeepan\S,~\IEEEmembership{Senior Member,~IEEE,} Karina Gomez\ddag,~\IEEEmembership{Member,~IEEE,} Laurent Reynaud\dag,~\IEEEmembership{Member,~IEEE,} Tinku Rasheed\ddag,~\IEEEmembership{Member,~IEEE.}
\IEEEcompsocitemizethanks{\IEEEcompsocthanksitem \S~Kandeepan Sithamparanathan is with the School of Electrical and Computer Engineering, RMIT University, Melbourne, Australia. E-mail: kandeepan@ieee.org.
                          \IEEEcompsocthanksitem \ddag~Karina Gomez and Tinku Rasheed are with CREATE-NET, Via Alla Cascata 56/C, 38123, Trento, Italy. E-mail: name.surname@create-net.org.
                          \IEEEcompsocthanksitem \dag~Laurent Reynaud is with the Orange Labs, France Telecom,  2 Avenue Pierre Marzin, 22307 Lannion, France. E-mail: laurent.reynaud@orange.com.
                          }
\thanks{A preliminary version of the mathematical model and the scenarios presented in this manuscript appeared in the proceedings of IEEE PIMRC'11~\cite{PIMRC}. In this manuscript, we have progressed substantially with the modeling aspects and also with respect to the implementation and simulations.}}

\markboth{To appear in IEEE Transactions on Aerospace and Electronic Systems}%
{Shell \MakeLowercase{\textit{et al.}}: Computer Society Journals}
%


\IEEEcompsoctitleabstractindextext{%
\begin{abstract}

Hybrid aerial-terrestrial communication networks based on Low Altitude Platforms (LAPs) are expected to optimally meet the urgent communication needs of emergency relief and recovery operations for tackling large scale natural disasters. The energy-efficient operation of such networks is important given the fact that the entire network infrastructure, including the battery operated ground terminals, exhibits requirements to operate under power-constrained situations. In this paper, we discuss the design and evaluation of an adaptive cooperative scheme intended to extend the survivability of the battery operated aerial-terrestrial communication links. We propose and evaluate a real-time adaptive cooperative transmission strategy for dynamic selection between direct and cooperative links based on the channel conditions for improved energy efficiency. We show that the cooperation between mobile terrestrial terminals on the ground could improve the energy efficiency in the uplink depending on the temporal behavior of the terrestrial and the aerial uplink channels. The corresponding delay in having cooperative (relay-based) communications with relay selection is also addressed. The simulation analysis corroborates that the adaptive transmission technique improves the overall energy efficiency of the network whilst maintaining low latency enabling real time applications.


\end{abstract}

}

\maketitle

\IEEEdisplaynotcompsoctitleabstractindextext

%
\IEEEpeerreviewmaketitle

\section{Introduction}\label{sec:introduction}
In the context of emergency situations, relief response teams need to be able to rely on robust and reliable communications. This observation can become a significant issue when the occurrence of a disaster leaves the existing network infrastructure in a state of partial or complete failure and therefore unable to meet the first rescuers' requirements. This recently triggered renewed research interest for reliable emergency communication techniques, thereby emphasizing the challenges embodied by network properties such as rapid deployment, self sustainability and interoperability as well as flexibility and scalability~\cite{WPMC2011}\cite{pe-wasun}.

In this regard, aerial telecommunication platforms and in particular Low Altitude Platforms (LAPs) have increasingly been the focus of researchers' and entrepreneurs' attention~\cite{Deaton}\cite{HAP_Book1}. These platforms are increasingly believed to offer the potentiality to effectively complement conventional satellite or terrestrial telecommunication infrastructures, for example as announced by Google~\cite{google}. These platforms are in fact thought to possess instrumental abilities for the support of efficient emergency communications, able to sustain stringent network coverage and capacity demands. Firstly, aerial vehicles, which can take the form of many types of aerodynes (airfoil-based crafts) or aerostats (lighter-than-air crafts) with various mobility patterns, can access disaster areas with a relative ease, often regardless of the state of the terrestrial transportation infrastructure. Moreover, these aircrafts feature fast deployments, multi-purpose operations at varying altitudes in the range of a few dozen to a few thousand meters, as well as modular communication payloads. All these abilities, combined with their aptitude to support frequent Line Of Sight (LOS) with terrestrial equipment and end-user devices, explain the uptake of LAPs for the support of Public Protection for Disaster Relief (PPDR) communication systems.

However, it is also generally accepted that unleashing the greater potential of aerial networks for emergency communications widely depends on how efficiently airborne platforms are integrated with satellite and terrestrial telecommunication infrastructures \cite{Deaton,TAES1,TAES2,TAES3}. Satellite-based emergency recovery solutions, which by nature are resilient against large-scale catastrophes, have been thoroughly investigated \cite{del_re}. Besides, their most noticeable shortcomings, such as the hardware complexity and limited data service support found on satellite phones as well as the relatively long deployment and configuration time of Very Small Aperture Terminal (VSAT) solutions are largely averted when used in complementarity with aerial networks. Integrated aerial-satellite infrastructure can indeed provide standard end-user devices with adapted and scalable access network capacity and coverage thanks to low delay and high capacity aerial links, while reliable backhauling links to remote networks (e.g. PPDR headquarters) can be supplied by the satellite segment.

Likewise, emergency communications can greatly benefit from an integration of aerial and terrestrial networks, enabling potential interoperability with the undamaged or repaired terrestrial equipment on the disaster zone and thus increasing the diversity of access and backhaul links. Furthermore, another strong interest of hybrid aerial-terrestrial integration is highlighted by the frequent case when dedicated telecommunication ground equipment is deployed on carefully chosen spots of the considered emergency area to complement other communication network segments or to bring specific functionalities~\cite{ABSOLUTE} including the support of broadband satellite links and interoperability with legacy PPDR systems such as TErrestrial Trunked RAdio (TETRA) communications.

In any case, an emergency network, including the hybrid aerial-terrestrial communication architecture which is studied in this work, faces stiff constraints regarding its overall energy consumption. In fact, disaster areas are likely to experience power outages and the whole emergency communication network has therefore to demonstrate its ability to ensure energy-efficient operation so as to maximize its own survivability. In particular, both terrestrial and aerial emergency network segments must meet energy efficiency challenges, as both ground equipments and LAPs frequently need to operate on battery power, without permanent access to a reliable energy source. Moreover, since the significant energy spent in the ground-to-LAP communication links highly depends on the availability and the quality of the related uplink channels, the consumption pattern of the communication links between the ground equipment and the LAPs must be carefully investigated and optimized. In this context, multiple energy saving methods, among which the sleep-wakeup strategy, have been considered for energy-constrained wireless mobile architecture~\cite{FP7_EARTH}. Yet, ground-to-LAP link performance in terms of delay and throughput cannot be conveniently compromised and consequently, those energy-efficient schemes cannot be easily transposed into the context of emergency and public safety communications.

In this work, we investigate the design and evaluation of an adaptive cooperative (relay-based) communication scheme intended to extend the survivability of the terrestrial nodes in an aerial-terrestrial network. To this end, we refine and significantly extend several of our previous studies~\cite{PSATS11, PIMRC, E2NETS12}: initially we have presented a preliminary analysis for the hybrid aerial-terrestrial systems for an uplink Rayleigh fading channel, which we later supplement with an analytical result for an uplink Ricean channel. By highlighting that the Ricean model is more realistic for the ground-to-LAP communications with LOS conditions, we outline in this paper:

\begin{itemize}

\item The design of an adaptive transmission scheme at the terrestrial nodes that selects either the direct or the cooperative relay link in order to minimize transmission power constrained by the bit error probability.

\item The delay analysis due to cooperation (relaying information) by considering two classes of traffic at a relay node, the local and relay traffics being subsequently allocated different priorities.
\end{itemize}

The rest of the paper is organized as follows. In Section~\ref{sec:architecture}, we present the system model and scenario description for the rapidly deployable communication network architecture for emergency response situations. A brief description of related works and current progress with aerial telecommunications network are provided in Section~\ref{sec:related}. In Section~\ref{sec:models}, we discuss the communication model and related assumptions. In Section~\ref{sec:Energy-Efficiency}, we analyze the energy efficiency aspects of cooperative communications in the aerial-terrestrial network segment and model the power efficiency for single source-multiple relay and multiple source-multiple relay scenarios. In Section~\ref{sec:algorithms}, the adaptive algorithm for relay node selection for establishing energy-efficient communication links in aerial-terrestrial networks is presented. Section~\ref{sec:analysis} discusses the evaluation of energy efficiency of communication links in the aerial-terrestrial deployment and finally Section~\ref{sec:conclusion} provides some concluding remarks.

\section{System Model and Scenarios}\label{sec:architecture}
In this section, we present the hybrid LAP-terrestrial system environment mainly considering a disaster recovery scenario, derived from the research project that we currently work on~\cite{ABSOLUTE}. To begin with, we present the global architecture with a combination of aerial, terrestrial and satellite communication links with the aim to maximize network availability and allow rapid and incremental network deployment. Such scalable and flexible network with adequate levels of mobility support and energy efficiency is suited for large disaster recovery operations~\cite{pe-wasun}. Finally, we present the sub-system architecture that we consider in this work addressing cooperation amongst the ground terminals for energy-efficient transmissions.

\subsection{The Global Architecture for Aerial-Terrestrial Systems}

\begin{figure*}
\begin{center}
\includegraphics[width=0.90\textwidth, height=75mm]{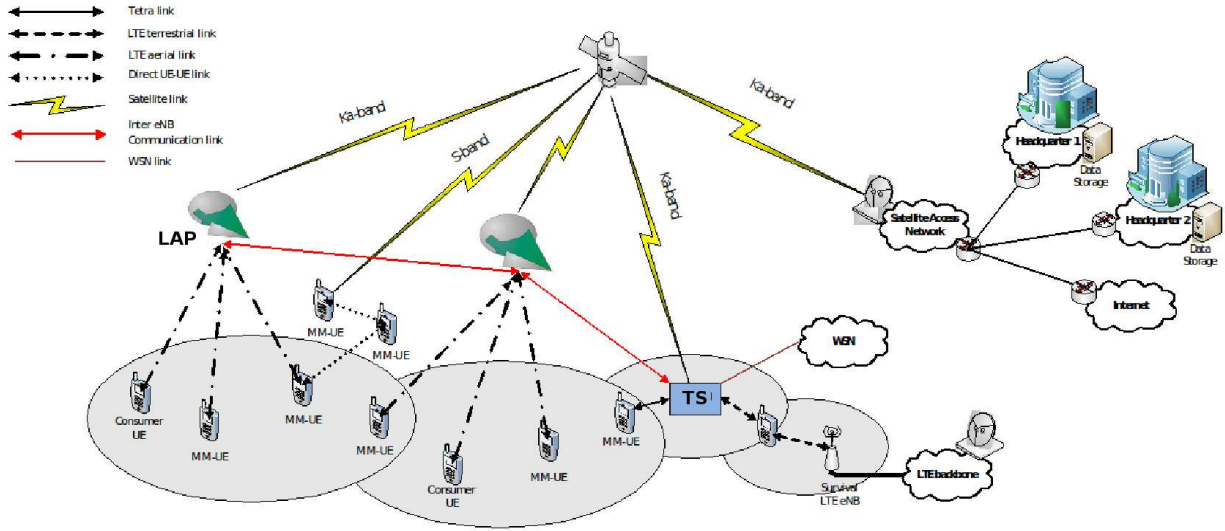}
\caption{The global architecture for the heterogeneous LAP-Terrestrial communication network with opportunistic communication links: In our work presented in this paper though, we consider a single LAP system only, without the satellite backhaul \cite{ABSOLUTE}.}\vspace{-1.3em}
\label{fig2}
\end{center}
\end{figure*}

The global network architecture is sketched in Fig.~\ref{fig2} which shows an emergency communication environment managed by two LAP base stations, which are linked to multiple terrestrial UEs and gateway nodes supported by satellite backhaul links. As discussed before, depending on the particular requirement for a communications system to support emergency services in disaster-affected areas, LAP platforms may constitute the deployment of one or more aerial devices which carry communication payload (4G LTE) with the backhaul links supported by satellite communication (DVB-S2/RCS uplink broadcast). The first responders in the terrestrial segment are equipped with multi-radio mobile terminals (UE), with radio technologies such as LTE/WiMAX to communicate with the LAPs and WiFi/WPAN (IEEE 802.15.4) to communicate with the ground terminals.

\subsection{Sub-System Architecture Considered for Energy-Efficient Transmissions}
In our work on energy-efficient transmissions for aerial-terrestrial systems presented here, we mainly consider the sub-system corresponding to a single LAP base station with many UEs, which is a part of the global architecture presented in Fig.\ref{fig2}. From Fig.\ref{fig2}, we can notice that for the coverage region under one LAP base station, a particular UE, in an event of battery depletion or a bad channel (e.g., due to obstacles), could use another UE as a relay in communicating with the LAP station. Alternately a UE could utilize the terrestrial gateway (TG) in order to communicate with the LAP. In the network model, we assume context-awareness, meaning that the terrestrial nodes are assumed to be aware of the necessary transmission and channel parameters such as channel gains etc. of the neighboring nodes. The reporting within the network for context-aware learning is discussed separately in the next section.

Moreover, cooperation between the ground terminals is also considered viable through the multi-radio interfaces. We also assume that all the ground terminals have access to the communication from the LAP platform in the downlink. For the uplink however, the communication link between the ground terminal and the LAP platform is more energy-constrained due to power scarcity in the ground terminals considering the higher transmission power required to reach the LAP. Such communication need to be optimized in terms of energy consumption as the uplink terrestrial-LAP channel gains can vary due to various conditions depending on the geographic area.

\section{Related Work}\label{sec:related}
Here, we present an overview of the most relevant related works on (i) energy-efficient cooperative-relay techniques in wireless communications and (ii) solutions for aerial-terrestrial network deployments during emergency scenarios.

\subsection{Relay Selection in Wireless Communication}

Cooperative wireless communication schemes, in which wireless nodes cooperate with each other in transmitting information, promise significant gains in overall throughput and improvement in terms of coverage and reliability. However, cooperation schemes can be also utilized for improving the energy efficiency of the whole network. In \cite{PTrading} authors perform relay selection for trading power in wireless communications. In~\cite{relay2} spectrally-efficient relay selection techniques with limited feedback are proposed but the authors did not consider energy efficiency related problems. Similar work in~\cite{relay5} addresses relay selection mechanisms, where a decentralized, opportunistic relaying scheme is presented to select the best relay node based on instantaneous end-to-end channel conditions and statistics. In this scheme the best relay node is selected from a set of available relay nodes in order to be used for cooperative communication between the source and the destination nodes based on the channels for improved capacity.
Relay selection schemes taking into account the energy consumption of the system are presented in~\cite{relay6,new_ref1,new_ref2,relay4,relay1}. In~\cite{relay6}, the authors consider a cluster of nodes cooperating together and analyze the corresponding energy efficiency due to cooperation. Madan et al.~\cite{relay4} consider selecting relays using channel state information for energy efficiency, the relays here are selected taking into account the channel state information (CSI) of the direct and cooperative paths. The authors in~\cite{relay1} propose a cooperative strategy for uplink and downlink communications by considering the location of the relay nodes in the environment and selecting the best relay which will result in an overall minimum energy consumption. In all these works, optimal power allocation is not considered in the cooperative communication link which we present in our work together with an adaptive link selection algorithm. Furthermore, we also analyze the delay associated with cooperation considering two classes of traffic at a relay node namely the local traffic and the relay traffic. The power controlling techniques for energy efficiency discussed in \cite{new_ref1} and \cite{new_ref2} are based on different objectives than to what we consider in this paper, in the former the authors consider power controlling for maximizing the total capacity and in the latter the authors consider power controlling for minimizing outage. In our work however we consider power controlling for maintaining the bit error probability at the destination terminal as described in detail in the subsequent sections.

In the references \cite{PControl1} -\cite{PControl8} the authors provide an optimum power allocation mechanism for minimum outage or to maximize capacity or to maximize the average SNR or to minimize the overall bit error rate, which are very well different from addressing optimal power allocation for energy efficiency with a BER constraint which we treat in this paper, and moreover none of the above mentioned papers address adaptive relay selection which we present together with a proposed algorithm. In \cite{PControl9}, the authors address power allocation for saving power for a cognitive radio based relay system with primary and secondary user model which is different to our model in which all the radio nodes have equal rights to access the spectrum, again in \cite{PControl9} the authors do not provide any adaptive algorithm as presented in our paper.

\subsection{Aerial-Terrestrial Networks for Emergencies}

Historically, the satellite links are the classic aerial network used in case of emergencies due to its coverage advantage as described in~\cite{gps,gps1}. In~\cite{gps} a proposal for transitioning from terrestrial-based navigation seeks to implement satellite and airborne surveillance operations as the primary navigation means is presented. The authors discuss about the installation and use of modern navigation and surveillance equipment which is mandated by the regulatory organizations. Further, a framework of the first phase of a post-disaster emergency situation is introduced in~\cite{gps1}. The purpose of the paper is to study a solution allowing to establish very
quickly a minimal low bit rate satellite link in Ku/Ka or Q/V bands for emergency communications. The paper also discusses the security and cost saving benefits drawn from the use of aerial networks, in comparison to terrestrial network in the context of emergencies situation.

In recent years, aerial-terrestrial network deployment using Aerial Platforms (AP) has been studied for the provision of wireless communications and related services~\cite{HAP_Book1,HAP_Book2}, \cite{kandee1}-\cite{kandee6}. Several projects have also investigated the communications aspects for APs~\cite{Project1,Project2}. The conclusions of these works are that APs have a number of potential advantages over conventional technologies, including (i) high capacity regional coverage, (ii) rapid deployment and (iii) potentially low cost. In~\cite{PACE_ALOI}, authors present an overview of how space technologies and satellite applications can mitigate the impact of natural and man-made disasters providing the reader with an overview of the future telecommunications architectures for both monitoring and disaster recovery purposes. The main features of high altitude platforms (HAPs) in terms of rapid deployment and broadband coverage are discussed here. The authors also point out that many issues and research aspects regarding the possible integration of different architectures and technologies are still open questions and require a deeper investigation in the future. These questions are partially addressed in~\cite{HAP_energy}. In this paper, the authors investigate schemes of the HAP-based emergency communications network to support emergency calls and alarming services in case of large scale disaster in mountainous area. The proposed network architecture consists of a two-hop relay system based on IEEE 802.16j standard (WiMAX). Two network scenarios are deeply analyzed by the authors according to the different roles that the HAP plays in the relay system. The first network scenario is an integrated satellite HAP-terrestrial system and the second one is related to an integrated HAP-terrestrial system. In order to explore the use of cellular technologies in HAPs, an analysis of the link budgets for high altitude airborne cellular base stations is presented in~\cite{HAP_antenna}. The authors developed a two-dimensional aircraft simulation environment in order to evaluate the optimal link budget of an antenna system coupled with a Code Division Multiple Access (CDMA) base station positioned in different altitudes. The presented results constitute a basis for low-risk implementation for a high altitude airborne deployable cellular base station.

The authors in~\cite{Hybrid} investigate how the HAP can effectively support multimedia broadcast/multicast service (MBMS) in a scenario where the existing terrestrial network infrastructure is not available. They propose radio resource management (RRM) techniques to be implemented in a multi-hop scenario in which MANETs cooperate with a HAP system to reach the common goal of enhancing the access to MBMS services. This work shows how a cooperative behavior of HAP and MANET systems can provide coverage to users in an incident area. A similar study in~\cite{Indonesia_ATN} proposes a LAP platform consisting of tethered balloons combined with IEEE 802.11 technology in order to build an ad-hoc communication network backbone supporting emergency medical services in natural disaster areas where the telecommunications facility has been seriously damaged.

Recently, the Unmanned Aerial Vehicles (UAVs) are being also investigated for supporting emergency communications. In~\cite{uav1} an embedded hardware/software
architecture specifically designed to be applied on mini/micro UAV to operate in dangerous-dirty-dull situations is presented. The proposed UAV architectonic proposal involves the following key elements: a LAN-based distributed and scalable hardware architecture, a service/subscription based software architecture, and an abstraction communication layer. Similar concepts are also discussed in~\cite{uav2}, where an experimental UAV system for search-and-rescue challenge is presented. The proposed UAV robot consists of a standard hobby remote-control airframe that is modified for autonomous flight, GPS-based navigation, ground image acquisition, and payload delivery. These components are integrated in order to provide capability for the search and rescue mission in a low-cost platform. In both papers~\cite{uav1,uav2}, the overall system architecture and system considerations are also discussed by the authors, who highlight the advantages of using aerial-terrestrial network deployments for emergency communications.

\subsection{Contributions}

The aforementioned studies however only investigate the deployment of aerial-terrestrial networks and their subsequent advantages during emergency situations without taking into account the energy aspects during transmissions. Since the base station is an aerial node with a high cost aerial link for all nodes compared to the terrestrial links and given that mobile users, LAPs and HAPs are battery-powered units, therefore the access cost to reach the aerial base station needs to be optimized. Energy-efficient protocols and algorithms are required to better implement such networks and extend their operations, a topic which we address in our paper here.

In this work, we have designed of an adaptive transmission scheme at the terrestrial that selects either the direct or the cooperative relay link in order to minimize transmission power-constrained by the bit error probability. The concept and mathematical models presented in this paper can be used as an input to design energy-efficiency clustering techniques and energy-efficiency cooperative communication schemes for out-of-coverage scenarios. Indeed, the difference in the communication channels between terrestrial nodes used for relaying, versus the channel to the LAPs are completely modelled in this work. Moreover, the delay analysis due to cooperation (relaying information) by considering two classes of traffic at a relay node namely the local and the relay traffic having different priorities is also considered and measured in this work. The proposed mechanism can be adapted to the regular cellular system with the base station on the ground however the performance gain may not be as significant as in the aerial based system. This is because the aerial base station is relatively far from the terrestrial nodes compared to the terrestrial base station, the treatment of this however is beyond the scope of this paper.

\section{Communication Model and Assumptions}\label{sec:models}
We present the cooperative networking model here by initially considering a single terrestrial source ($S$), multiple relays ($R$) and a single destination $D$ (which is the LAP) cooperative networking system and then extending it to a multiple-source and multiple-relay scenario with a single destination which is the LAP. Note that in our analysis, we consider only a single destination (LAP) considering the fact that the LAP could cover a vast geographic area to serve many terrestrial nodes. The analysis however could be extended for a multiple-LAP scenario with increased complexity. The terrestrial nodes are assumed to be moving slowly giving rise to a slow fading scenario, and furthermore we assume no multipath signals resulting in a flat fading channel.

\subsection{Single-Source, Multiple-relay Model}
\begin{figure}
\begin{center}
\includegraphics[width=0.45\textwidth]{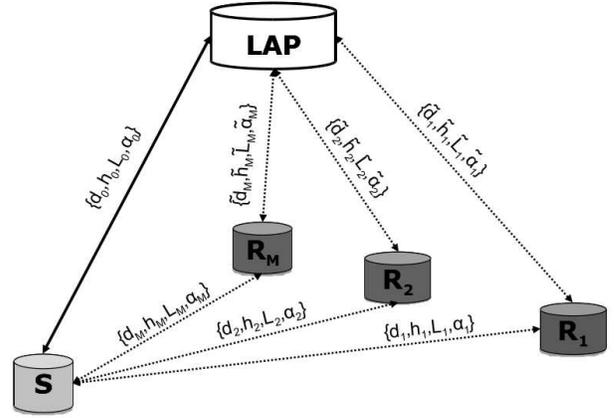}
\caption{Scenario for Hybrid LAP-Terrestrial communications with cooperative relay links}
\label{fig3}
\end{center}
\end{figure}

The communication model for the uplink terrestrial-LAP channel and the terrestrial channel models for a single terrestrial source, a cooperative networking system made of multiple relays and a single destination (LAP) is presented here. The channel models and the related parameters for the hybrid LAP-terrestrial system is depicted in Fig.~\ref{fig3}. In our model:

\begin{itemize}
\item $h_i$ are the small scale slow flat fading gains,
\item $\alpha_i$ are the pathloss exponents,
\item $d_i$ are the S-R distances,
\item $L_i$ are the mean pathloss with $i \in \{1,2 \dots M\}$ for the source to relay links ($S \rightarrow R_i$) as shown in Fig.~\ref{fig3}.
\end{itemize}

For the links corresponding to the relays to the LAP ($R_i \rightarrow LAP$) the respective parameters are given by $\tilde{h}_i, \tilde{\alpha}_i, \tilde{d}_i$ and $\tilde{L}_i$, as depicted in the Fig.~\ref{fig3}. For the uplink channels from the relays to the LAP ($R_i \rightarrow LAP$) and from source to the LAP ($S \rightarrow LAP$) we consider Rice small-scale fading channel models~\cite{Rapp}, the corresponding channel gains vary over time based on the spatial separation between the nodes. For the terrestrial radios, we consider a Rayleigh channel model.

The mean pathloss for the wireless channels are defined by the path loss exponents $\alpha_i$ and $\tilde{\alpha}_i$ as described subsequently. Note that in the subsequent description, we ignore the subscripts for the parameters as they are true for all the channels. The mean pathloss is given by
\begin{eqnarray}\label{eq1}
L(d) = L(\bar{d})\left(\frac{d}{\bar{d}}\right)^{\alpha}
\end{eqnarray}
where $L(\bar{d})$ is the mean pathloss at a reference distance $\bar{d}$ which is given by the free-space pathloss~\cite{Rapp} equal to $L(\bar{d}) = \left(4\pi f \bar{d}/c\right)^2$, where $c=3\times10^8$ $m/s$ is the speed of light and $f$ is the carrier frequency. The mean channel power gains are given by
\begin{eqnarray}\label{eq2}
\gamma = \frac{1}{t}\int_0^t h^2(t)dt
\end{eqnarray}
Note that for the Ricean uplink channels, the channel gains contain both the direct-path and the multi-path component signal gains~\cite{Rapp}. Moreover, for the ease of analysis, we consider binary phase-shift keying modulation for the terrestrial links and differential phase-shift keying for the terrestrial-LAP uplinks, the corresponding received signals for the communication links described in Fig.~\ref{fig3} can be expressed in general as
\begin{eqnarray}\label{eq3}
r(t) = \frac{1}{\sqrt{L(d)}}h(t)s(t) + \upsilon(t)
\end{eqnarray}
where $s(t)$ is the transmitted signal and $\upsilon(t)$ is the corresponding additive Gaussian noise at the receiver with a double-sided power spectral density of $N_0/2$, we consider the same power spectral densities for all the receivers for simplicity. The corresponding probability of error for the terrestrial link and the terrestrial-LAP links are respectively given by~\cite{Rapp},\\
For links $S \rightarrow R_i$
\begin{eqnarray}\label{eq4}
\Pi_1 &= 0.5(1 - \sqrt{\frac{\Gamma}{(1+\Gamma)}})
\end{eqnarray}
For links $R_i \rightarrow LAP$
\begin{eqnarray}\label{eq5}
\Pi_2 &= \frac{1+K}{2(1+K+\Gamma)}\exp(\frac{-K\Gamma}{\Gamma+1+K} )
\end{eqnarray}
where $K$ is the Rice factor~\cite{Rapp}, $\Gamma = E_b\gamma/N_0$ is the mean received SNR for the given link, $E_b$ is the received bit energy given by $E_b = P_tG_tG_r/[\Delta L(d)]$, $G_t$ and $G_r$ are the transmit and receive antenna gains, $P_t$ is the transmit signal power, and $\Delta$ (bits/s) is the data rate for the given link. The relaying protocol is based on decode and forward technique where a the relay node $R_i$ receives the signal $r_i(t)$ from $S$, detects it, and transmits the corresponding information to the LAP (destination). In the direct link the LAP directly receives the signal $r_0(t)$ from the source node $S$.

\subsection{Multiple-Source, Multiple-relay Model}
The multiple-source, multiple-relay and single-destination model is simply an extension of the single-source multiple relay scenario presented above. In the multiple-source case, we consider a total number of $K+1$ terrestrial nodes in the network that can act either as a source or a relay in the network such that every single-source node will have $K$ number of relays. When multiple nodes are acting as source nodes, the communication process related to every single-source node is given by the same model presented before for a single-source multiple-relay scenario, as depicted in Fig.~\ref{fig3}. Therefore, the multiple-source multiple-relay cooperative networking scenario is basically a combination of individual single-source multiple-relay communication processes.

\subsection{Context-Aware Reporting and Outdated Information}
Context-aware capabilities are required in order to know all the parameters used by the network model. In our scenario, these capabilities can be implemented easily through the use of gossip protocols since in the scenario the exchange of the context information is required only between nodes with a distance of one hop (see Fig.~\ref{fig3}). Many gossip protocols have been proposed for wireless ad hoc networks, and most of them are based on some variants of flooding with respect to the network density \cite{gossip,gossip1,gossip2}. Additionally, many gossip approaches have been studied, where only the nodes within a specific region broadcast or forward a message with a given probability, to reduce the overhead in the network \cite{gossip3,gossip4}. Additionally, this information can also be included in the overhead introduced by the upper layers (i.e. routing protocols or application overhead).

In our network model all the nodes periodically report contextual information by means of a single hop broadcast packet transmission. The overhead of the reporting packets is kept to a bare minimum to ensure no unwanted energy is consumed in the periodic reporting. The corresponding energy cost for the context aware reporting is also considered in our work as described later. In the periodic reporting every node transmits its address, geographic position (e.g. GPS co-ordinates), and the effective istropically radiated power ($EIRP = P_tG_t$) to every other node by means of broadcasting. Using the received information from the reporting packet every node will then make itself aware of the relative distance $d_i$ and the temporal pathloss between itself and that particular node which sent the report. The temporal pathloss, that includes both the mean pathloss $L_i$ and the temporal channel variation $\gamma_i$ together, is estimated by computing the received power level of the reporting packet and using the knowledge of the transmit $EIRP$ and the receiver antenna gain. Therefore, every node in the network becomes aware of the parameters such as $d_i$ and total channel gain $\gamma_i/L_i$ which are then used for the proposed cooperative relaying solution. Note that the knowledge about the radio environment such as the node positions and the neighboring nodes etc. can also be learnt by individual nodes using spectrum sensing techniques and learning techniques as discussed in the literature of cognitive radios \cite{kandee7}-\cite{kandee20}, in this paper however we do not treat such but rather assume perfect knowledge are known by means of sharing/reporting.

The selection of the reporting period $\tau_R$ is quite important to make sure the network does not operate with outdated information (i.e. due to the variation of the environment such as the wireless channel, position of a node etc.). We assume a slow fading channel with a coherence time of $T_{coh}$ \footnote{The time that the channel remains unchanged \cite{Rapp}.}, and assume $T_{coh} > \tau_R$ for having updated contextual information in the network. It is clearly seen that when the channel changes rapidly with $T_{coh} << \tau_R$ the network would have outdated network/channel information (note that when $T_{coh}< \tau_R$ there will be still strong correlation between the new channel gain and the outdated channel gain, and only when $T_{coh} << \tau_R$ we will see significant deviations in the channels gains). The treatment of such outdated network conditions is not considered in this paper but for future work. It is quite realistic to assume slow varying channels especially when the LAP terminal is stationary and the mobile terminals on the ground move at walking speeds, which is the prime use case scenario that we are interested in as described before. Moreover, dynamically adapting the reporting period $\tau_R$ such that it is kept less than $T_{coh}$ is also possible by tracking the channel variation using various channel tracking algorithms that are present in literature.

\section{Energy-efficient Cooperative Communications}\label{sec:Energy-Efficiency}
In order to better utilize the scarce battery power when communicating during live operations, we consider cooperative relaying communication techniques to minimize the energy consumption of the battery-operated terrestrial nodes. In order to communicate in an energy-efficient manner with the LAP, a source node will analyze its options to transmit via the direct link to the LAP and the relayed links through $K$ possible relay links in the network. In the subsequent section, we present how a source node could choose the most energy-efficient method to communicate with the LAP without compromising on the bit error rate to maintain a minimum level of service quality. In our subsequent analysis, we consider energy efficiency per unit time, leading us to power efficiency.

\subsection{Transmit Power Requirements for Single-Source and Single-Relay Network}
\begin{figure}
\begin{center}
\includegraphics[width=0.45\textwidth]{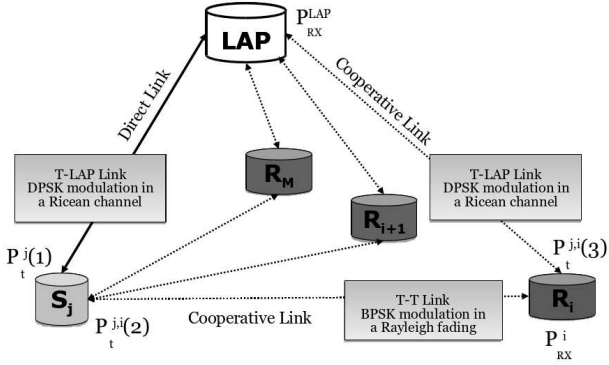}
\caption{Analytical model with single-source multiple-relay hybrid LAP-terrestrial communications.}
\label{fig4}
\end{center}
\end{figure}

For a given source node $S$, the minimum power requirement for the direct link with a bit error constraint of $\xi$ is analyzed here. For convenience, we consider a single-source and single-relay scenario at the beginning as depicted in Fig.~\ref{fig4}, and extend the scenario to a multiple-node case later. Moreover, we consider the communication parameters described in the previous section by introducing the subscripts $1, 2, $ and $3$, where subscript $1$ corresponds to the parameters for the direct link, subscript $2$ corresponds to the parameters for the relay link between $S$ and $R$, and subscript $3$ corresponds to the parameters related to the relay link between $R$ and $LAP$, as depicted in Fig.~\ref{fig4}.

By considering the bit error expression in ($5$) for the direct link, the transmit power requirement $P_t(1)$ at node $S$ can be iteratively computed using the Gradient Descent method for some $\epsilon_1 >0$ and $\Pi_2 = \xi$ given by,
\begin{eqnarray}\label{eq7}
\hat{P}_{t_{n+1}}(1) = \hat{P}_{t_{n}}(1) - \epsilon_1 \Lambda(\hat{P}_{t_n}(1))
\end{eqnarray}
where, $\Lambda(\hat{P}_{t_n}(1))$ (from equation (\ref{eq5})) is given by,
\begin{eqnarray}\label{eq7_1}
\Lambda(\hat{P}_{t_n}(1)) = \frac{d[\xi - \Pi_1(P_t(1))]}{dP_t(1)} = \frac{BV_1V_2[V_1K_1 + 1]}{[1+K_1 + \Gamma_1]} \nonumber\\
V_1 = \frac{1+K_1}{2[1+K_1+\Gamma_1]} \text{\ \ \, \ \ } V_2 = \exp\left(\frac{-K_1\Gamma_1}{1+K_1+\Gamma_1} \right)
\end{eqnarray}
and $B = \frac{G_1^t G_1^r \gamma_1}{L_1(d_1)N_0(1)\Delta_1}$, where as indicated before the subscript $1$ corresponds to all the parameters associated with the direct transmission link from $S \rightarrow LAP$. Therefore, for sufficient iterations (sufficient $n$) the minimum power requirement $\hat{P}(1)$ at the source node $S$ can be determined for the direct link.

The overall bit error probability for the $S\rightarrow R_i \rightarrow LAP$ communication based on the decode and forward cooperative strategy depends on the communications between $S \rightarrow R_i$ and $R_i \rightarrow LAP$. If $\Pi_1$ and $\Pi_2$ are the bit error probabilities respectively of the sequential transmission $S\rightarrow R \rightarrow LAP$ the overall error probability is then given by $\overline{\Pi} = \Pi_2(1-\Pi_1) + \Pi_1(1-\Pi_2)$. Since the probability values are small, $\overline{\Pi}$ can be approximated to $\overline{\Pi} \approx \Pi_1 + \Pi_2$. Then using (4) and (5) the overall bit error rate for the relay transmission is given by,
\begin{eqnarray}\label{eq6}
\overline{\Pi} = 0.5\left(1 - \sqrt{\frac{\Gamma_2}{(1+\Gamma_2)}}\right) + \nonumber\\
\frac{1+K_3}{2(1+K_3+\Gamma_3)}\times\exp\left(\frac{-K_3\Gamma_3}{\Gamma_3+1+K_3} \right)
\end{eqnarray}
Considering the above equation, the transmit power requirement $\{P_t(2), P_t(3) \}$ at the source and relay nodes respectively for a bit error constraint of $\xi$ is given by,
\begin{equation}\label{eq8}
P_t(2) = \frac{L_2(d_2)N_0(2) \Delta_2\lambda^2}{G^t_2G^r_2\gamma_2(1-\lambda^2)}
\end{equation}
where $\lambda$ is a function of the relay transmit power $P_t(3)$ given by,
\begin{eqnarray}\label{eq9}
\lambda &= (1-2\xi) + \frac{1+K_3}{(1+K_3+\Gamma_3)}\exp\left(\frac{-K_3\Gamma_3}{\Gamma_3+1+K_3} \right)
\end{eqnarray}
with $\Gamma_3 = \frac{G^t_3G^r_3\gamma_3P_t(3)}{ L_3(d_3)N_0(3) \Delta_3}$. Note that, as mentioned before, subscript $2$ corresponds to the parameters related to the link $S \rightarrow R$ and subscript $3$ corresponds to the parameters related to the link $R \rightarrow LAP$.

For the cooperative link, the optimum power allocations between the source and the relay nodes that would minimize the overall power consumption for the given bit error rate $\xi$ needs to be calculated for the source node to choose between the direct and cooperative links for power efficiency. Considering equation (\ref{eq9}), since $P_t(2)$ is function of $P_t(3)$, the optimum power allocation between $S$ and $R$ for the cooperative transmission can be found by,
\begin{equation}\label{eq11}
\{\hat{P}_t(2), \hat{P}_t(3) \} = \arg \min_{P_t(3)} \{P_t(2)+P_t(3)\}
\end{equation}
Note that $P_t(2)$ is a strictly decreasing function in the domain of $P_t(3)$ and given that $P_t(3)$ is a strictly increasing (linear and convex) function in its own domain the summation $X= P_t(2)+P_t(3)$ becomes convex. Since $X=P_t(2)+P_t(3)$ is a convex function in the domain of $P_t(3)$, the minimum of $X$ in $P_t(3)$ would provide the optimum power allocation $\{\hat{P}_t(2), \hat{P}_t(3) \}$ for the source and relay nodes. From equations (\ref{eq8}) and (\ref{eq9}), the first derivative of $X$ is given by,
\begin{equation}\label{eq12}
\frac{dX}{dP_t(3)} = 1+\frac{dP_t(2)}{dP_t(3)} = 1 - \frac{2\lambda A_0 A_1 U_1^2 U_2}{[1-\lambda^2]^2}
\end{equation}
where,
\begin{eqnarray}\label{eq13}
A_0 = \frac{L_2(d_2)N_0(2)\Delta_2}{G_2^t G_2^r \gamma_2} \text{\ \ , \ \ } A_1 = \frac{G_3^t G_3^r \gamma_3}{L_3(d_3)N_0(3)\Delta_3} \nonumber\\
U_1 = \frac{1+K_3}{[1+K_3+\Gamma_3]} \text{\ \ , \ \ } U_2 = \exp\left(\frac{-K_3\Gamma_3}{1+K_3+\Gamma_3}\right)
\end{eqnarray}
and $\lambda = (1-2\xi) + U_1 U_2$. The optimum relay power $\hat{P}_t(3)$ is then given by,
\begin{eqnarray}\label{eq14}
\frac{dX}{dP_t(3)}_{|_{P_t(3)=\hat{P}_t(3)}} = 1 -  \nonumber\\
\frac{2\lambda A_0A_1U_1U_2(U_1K_3+1) }{[1+K_3+\Gamma_3][1-\lambda^2]^2}
\frac{dX}{dP_t(3)}_{|_{P_t(3)=\hat{P}_t(3)}} = 0
\end{eqnarray}
Here, we consider again the Gradient Descent method to solve for $\hat{P}_t(3)$ from the above equation for some $\epsilon_2 >0$. The solution for $\hat{P}_t(3)$ is then given by,

\begin{equation}\label{eq15}
\hat{P}_{t_{n+1}}(3) = \hat{P}_{t_{n}}(3) - \epsilon_2 \frac{d^2X}{dP^2_t(3)}_{|_{P_t(3)=\hat{P}_{t_n}(3)}}
\end{equation}
Then, for sufficient iterations (sufficient $n$) the minimum power requirement $\hat{P}(3)$ at the relay node $R$ can be determined for the cooperative link, and the optimum transmit power $\hat{P}_t(2)$ at $S$ is given by substituting $\hat{P}_t(3)$ in equation (\ref{eq8}) to obtain $\{\hat{P}_t(2), \hat{P}_t(3) \}$ for minimum power consumption in the cooperative link.

\subsection{Energy Efficiency in Multiple-Source Multiple-Relay Network and Link Selection}
The above given minimum power requirement analysis can be extended to a multiple-source, multiple-relay scenario. For the $j^{th}$ source node $S_j$ for $j = 1, 2 \dots M+1$ and the $i^{th}$ relay node for $i = 1, 2 \dots M$ \footnote{The source nodes and the relay nodes are indexed separately and independently to each other.}, let us define the (power/energy) efficiency factor $\beta_{j,i}$ to compare the energy efficiency between the direct and the relay links, given by,

\begin{equation}\label{eq10}
\beta_{j,i} \triangleq \frac{e_{j,i} (T)}{e^D_{j}(T)}
\end{equation}

where,
\begin{itemize}
\item $e_{j,i}$ is the total energy cost for relay link for source $j$ and relay $i$ over time $T$,
\item $e^D_{j}$ is the total energy cost for direct link for source $j$ over time $T$,
\end{itemize}

Based on (\ref{eq10}) for a given source node $S_j$ when $\beta_{j,i}>1, \forall i$ the direct link becomes more power-efficient than all the cooperative links in which case $S_j$ selects the direct link for its transmissions. When $\beta_{j,i}<1$ for some $i$ the corresponding relay link(s) become(s) more power-efficient than the direct link in which case $S_j$ selects the most energy-efficient cooperative relay link, and the corresponding power efficiency for source node $S_j$ is given by,
\begin{equation}\label{eq11}
\hat{\beta}_{j} = \min_i \{\beta_{j,i}\} \text{\ \ for $\beta_{j,i}<1$ }
\end{equation}
Note that when $\beta_{j,i}=1$ both relay and direct links become equally energy-efficient and it is desirable to use the direct link to minimize delays. In summary, the source node selects the most power-efficient link by choosing between the direct link and the relay links, the link selection is made adaptive in time considering the channel variation in time.

\subsection{Energy Consumption}\label{energy_comp}
The energy consumption due to transmissions over a given time period $T \in \mathbb{R}^+$ is considered to be the key performance metric here. The total energy consumption for the direct link $e_j^D$ and the relay link $e_{j,i}$ related to a particular source node $j$ and a relay node $i$ are respectively given by,
\begin{equation}\label{eq12}
e_j^D = P^{j}_t(1)T_{d} + P^{LAP}_{Rx}T_{d}
\end{equation}
\begin{equation}\label{eq13}
e_{j,i} = P^{j,i}_t(2)T_{d}+P^{j,i}_t(3)T_{d}+P^{i}_{Rx}T_{d}+ P^{LAP}_{Rx}T_{d}+E_{com}
\end{equation}
where,

\begin{itemize}
\item $T_{d}$ is the total time taken for data transmission in a period of $T$,
\item $P_t(1), P_t(2)$ and $P_t(3)$ are the transmit powers as described in Section-V-A,
\item $P^i_{Rx}$ is the required power to receive data at a given relay node ($R_i$),
\item $P^{LAP}_{Rx}$ is the required power to receive data at the LAP,
\item $E_{com}$ is the energy consumption at the transmitters (source and relay) to perform computations of optimum power controlling as descried in the previous section, which is assumed to be the same for all the terrestrial nodes.
\end{itemize}

Note that $P^i_{Rx}$ and $P^{LAP}_{Rx}$ both incorporate all the RF related hardware and baseband power consumption corresponding to receiving the signal, and in practice it is reasonable to assume that $P^i_{Rx} = P_{Rx}$ to be the same for all the terrestrial nodes. Moreover, the overhead energy that is required for context-aware reporting $E_{oh}$ is not included in the cooperative link energy cost for the $e_{j,i}$ given in (\ref{eq13}) but instead we include this cost in the total network energy cost considering the entire network operating with cooperative links. The total network energy cost for the cooperative/relay based multiple source multiple relay network $e_R$ is then given by,
\begin{equation}\label{eq13_1}
e^R = \sum_{j=1}^L e_{j,i} + E_{oh}
\end{equation}
where $L \leq K+1$ is the total number of source nodes that perform transmissions over a time period of $T$ seconds, and $E_{oh}$ is the total energy consumption for the context-aware reporting for the entire network considering both transmitting and receiving energy costs of the reporting packets. The two additional energy costs $E_{com}$ and $E_{oh}$ for the cooperation of nodes in the network are further described in Appendix-A.

The energy consumption for the adaptive link selection-based transmission strategy for a particular source node $j$ is given by,
\begin{equation}\label{eq14}
e_j = \min \{ e_j^D, e_{j,i}\} \text{\ \ \ } \forall i
\end{equation}
and therefore the total network energy consumption for the adaptive link selection-based transmissions for the multiple source multiple relay network is given by,
\begin{equation}\label{eq15}
e^A = \sum_{j=1}^L e_j + e_{oh}
\end{equation}
The energy costs for the cooperative systems given by $e^R$ and $e^A$ above are compared with the total energy consumption $e^D$ for the direct transmissions only case which given by,
\begin{equation}\label{eq15_1}
e^D = \sum_{j=1}^L e^D_j
\end{equation}

\subsection{Cooperation Delay in Multiple-Source and Multiple-Relay Network}
In the above presented analysis, we see how energy efficiency can be achieved considering multiple cooperative links for a given source node $S_j$. The energy efficiency however comes at a price of additional delay during cooperation when the relay nodes are acting as source nodes themselves. In such situations, the relay links give higher priority to their own data to be transmitted whilst delaying the data from other nodes to be relayed to the LAP. The corresponding analysis for the delay can be presented by considering a single-server queueing model. The corresponding analysis however is beyond the scope of this paper and we only present simulation results related to the delay analysis in later sections.

\section{Adaptive Energy-efficient Link Selection Approach}\label{sec:algorithms}
We present the procedure to implement the adaptive algorithm for the adaptive link selection-based transmission presented in the previous sections. Based on our approach, a source node $S_j$, in the presence of multiple communication paths to reach the LAP can choose the most energy-efficient link to communicate with the LAP depending on the energy efficiency factor $\beta_j$. The notation used throughout this section is shown in Table~\ref{tbl:notation}.

\begin{table}[t]
\renewcommand{\arraystretch}{1.4}
\centering
\caption{Notation}
\label{tbl:notation}
\begin{tabular}{r||l}\hline
\textbf{Notation}            &\textbf{Description}   \\
\hline
\hline
$S_j$             & The source node for $j = 1, 2 \dots K+1$\\
$R_i$             & The relay node for $i = 1, 2 \dots K$\\
$\beta_{j,i}$     & The energy efficiency factor for a $S_j$ with $R_{i}$\\
$P^i_{Rx}$        & The power requirement to receive data at \\
                  & a given relay node $R_i$\\
$P^{LAP}_{Rx}$    & The power requirement to receive data at the LAP\\
$P^{j}_t(1)$      & The transmission power for the aerial interface in $S_j$\\
$P^{j,i}_t(2)$    & The transmission power for the Terrestrial interface in \\
                  & $S_j$ to the cooperative link with $R_i$\\
$P^{j,i}_t(3)$    & The transmission power for the aerial interface in \\
                  & $R_i$ for the data received from $S_j$\\
$SNR^j_{1}$       & The SNR for the direct link $S_j \rightarrow LAP$\\
$SNR^{j,i}_{2}$   & The SNR for the cooperative link $S_j \rightarrow R_i$\\
$SNR^i_{3}$       & The SNR for the cooperative link $R_i \rightarrow LAP$\\
$E_{com}$         & Computational energy cost \\
\hline
\end{tabular}
\end{table}

\subsection{Procedure at the Source $S_j$}

Alg.~\ref{alg:link} describes the pseudocode for the energy-efficient link selection approach at the source node $S_j$ using multiple-relay nodes $R_i$. The algorithm is run at the source node $S_j$ and (i) it calculates the optimal power allocation for each cooperative link in the network and (ii) it selects either the direct link or the more energy-efficient cooperative link depending on the energy efficiency factor $\beta_j$ in order to communicate with the LAP (see also Fig.~\ref{fig4}). When the source node $S_j$ has a packet to be transmitted to the LAP, the source node $S_j$ performs the following steps:

\begin{itemize}

\item[a)] The source node $S_j$ calculates the bit error rate $BER_j$ for the direct link $S_j \rightarrow LAP$ using the transmission power $P^{j}_t(1)$ of the aerial interface and the channel conditions of the direct link $SNR^j_{1}$ ($S_j \rightarrow LAP$).

\item[b)] In order to minimize the overall power consumption, the source node $S_j$ calculates the optimum power allocation $P^{j,i}_{t}(3)$ for aerial interface of each relay node $R_i$ using the $BER_j$ for the direct link $S_j \rightarrow LAP$ and the channel conditions of the cooperative link $SNR^{j,i}_{2}$ and $SNR^i_{3}$ ($S_j \rightarrow R_i\rightarrow LAP$).

\item[c)] The source node $S_j$ calculates the optimum power allocation $P^{j,i}_{t}(2)$ for the Terrestrial interface of the source node $S_j$ according to the respective relay node $R_i$ and against the channel conditions of each cooperative link $SNR^{j,i}_{2}$ and $SNR^i_{3}$ ($S_j \rightarrow R_i\rightarrow LAP$).

\item[d)] $S_j$ calculates $\beta_{j,i}$ for each cooperative link in the network ($S_j \rightarrow R_i\rightarrow LAP$) using the transmission power $P^{j}_t(1)$ for the direct link, the transmission power $P^{j,i}_t(2)$ and $P^{j,i}_t(3)$ for each cooperative link and the power requirements $P^{i}_{Rx}$ and $P^{LAP}_{Rx}$ in order to receive data at relay node $R_i$ and LAP, respectively.

\item[e)] Finally, $S_j$ chooses between the direct link and the different cooperative link for power efficiency. It verifies  if $\exists\hat{\beta_{j}}$ where $\hat{\beta_{j}} = \min_i \beta_{j,i}$  for  $\beta_{j,i}<1$.

\begin{itemize}
\item[(i)] {\bf If $\hat{\beta_{j}}$ does not exist}, the source node $S_j$ sets the aerial interface equal to transmission power $P^{j}_{t}(1)$ and sends out the packet through the direct link to the LAP ($S_j \rightarrow LAP$).

\item[(ii)] {\bf If $\hat{\beta_{j}}$ exists}, the source node $S_j$ sets the Terrestrial interface equal to transmission power $P^{j,i}_{t}(2)$ and sends out the packet through the cooperative link to the LAP ($S_j \rightarrow R_i \rightarrow LAP$).

\end{itemize}
\end{itemize}

\begin{algorithm}[t]
\small
\caption{Adaptive Algorithm for the source node $S_j$ using multiple-relay nodes $R_i$.}
\label{alg:link}
\begin{algorithmic}[1]
\STATE {\bf Procedure at the source $S_j$ }
\IF {$Queue$ is not empty}
		\STATE $BER_j$ = calculateBER[$P^{j}_t(1)$,$SNR^j_{1}$]
		\FORALL{$R_{i}$ \text{  with  } $i=1, 2 \dots K $}
		\STATE $P^{j,i}_{t}(3)$ = calculatePt[$BER_j$,$SNR^{j,i}_{2}$,$SNR^i_{3}$]
		\STATE $P^{j,i}_{t}(2)$ = calculatePt[$P^{j,i}_{t}(3)$,$SNR^{j,i}_{2}$,$SNR^i_{3}$]
		\STATE $\beta_{j,i}$ = calculateBeta[$P^{j,i}_t(2)$,$P^{j,i}_t(3)$,$P^{j}_t(1)$,$P^{i}_{Rx}$,$P^{LAP}_{Rx},E_{com}$]
		\ENDFOR
		\IF{$\exists\hat{\beta_{j}} \rightarrow \hat{\beta_{j}} = \min_i \beta_{j,i}$ \text{ for } $\beta_{j,i}<1$ }
		\STATE $R$ $\rightarrow$ $R_{i}$
		\STATE $T_x$ = $P^{j,i}_{t}(2)$
		\STATE $p$ = Queue.next()
		\STATE sendCooperativeLink($p$)
		\ELSE
		\STATE $T_x$ = $P^{j}_t(1)$
		\STATE $p$ = Queue.next()
		\STATE sendDirectLink($p$)
		\ENDIF
\ENDIF
\end{algorithmic}
\end{algorithm}

\subsection{Procedure at the Relay $R_i$}

Alg.~\ref{alg:link2} describes the pseudocode for the energy-efficient link selection scheme implemented at the relay node $R_i$ using multiple-source nodes $S_j$. The algorithm is performed by the relay node $R_i$ in order to allocate the optimal transmission power to the aerial interface according to the data coming from a specific source node $S_j$. When the relay node $R_i$ receives a packet to be relayed to the LAP from the source node $S_j$, the relay node $R_i$ performs the following steps:

\begin{itemize}
\item[a)] The relay node $R_i$ sets the transmission power of the aerial interface equal to transmission power $P^{j,i}_{t}(3)$,
\item[b)] The relay node $R_i$ sends out the packet through the direct link to the LAP ($R_i \rightarrow LAP$).
\end{itemize}

\begin{algorithm}[t]
\small
\caption{Adaptive Algorithm at the relay node $R_i$ using multiple-source nodes $S_j$.}
\label{alg:link2}
\begin{algorithmic}[1]
\STATE {\bf Procedure at the relay $R_i$ }
\IF {receive($p$)=relay packet}
	\STATE $S$ $\rightarrow$ $S_{j}$
	\STATE $T_x$ = $P^{j,i}_{t}(3)$
	\STATE sendDirectLink($p$)
\ENDIF
\end{algorithmic}
\end{algorithm}

\begin{figure}
\begin{center}
\includegraphics[width=0.45\textwidth, height=50mm]{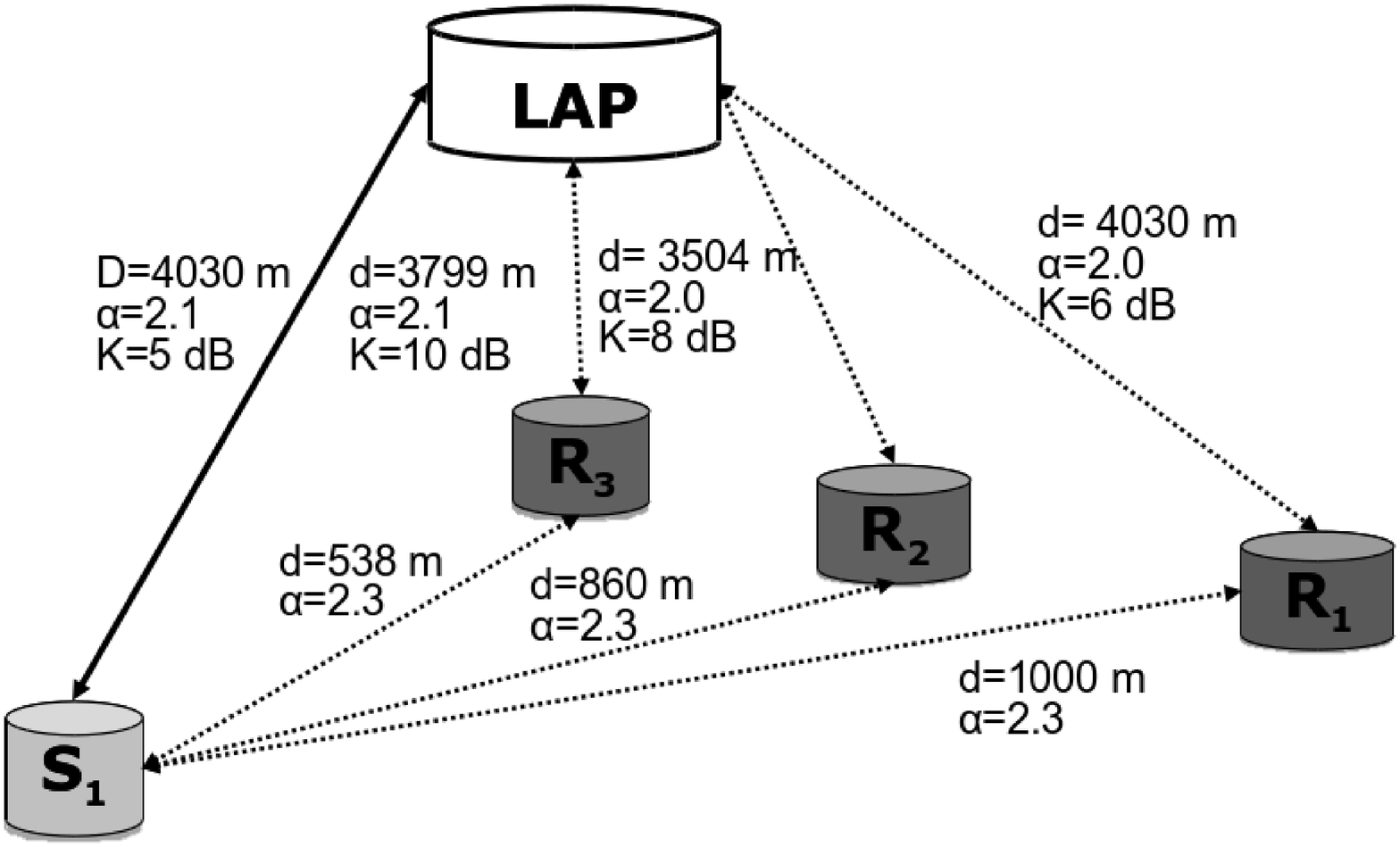}
\caption{Scenario for single-source multiple-relay hybrid LAP-terrestrial communications.}
\label{fig5}
\end{center}
\end{figure}

\begin{figure*}
\begin{center}
\includegraphics[width=0.95\textwidth, height=50mm]{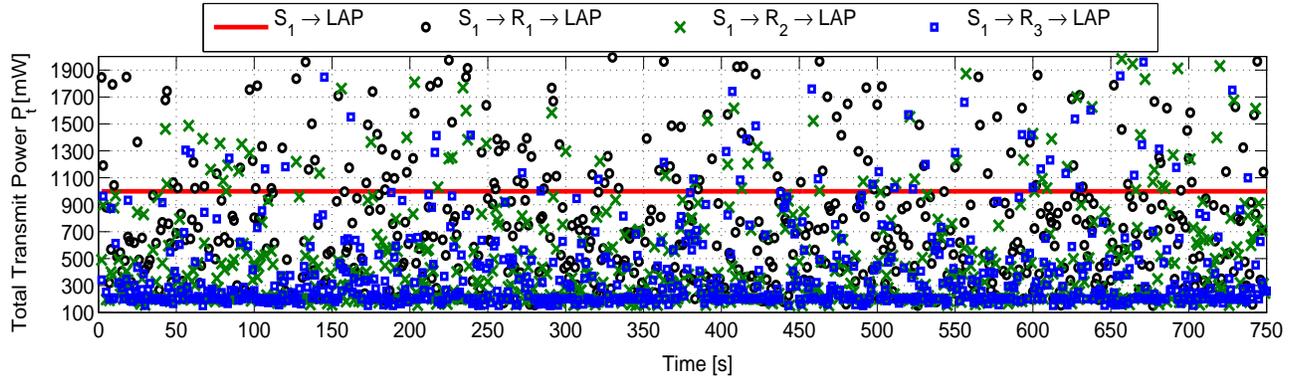}
\caption{Total transmission power calculated by the direct and cooperative links, before applying the adaptive scheme Vs. time for single-source multiple-relay hybrid LAP-terrestrial communications.} \vspace{-1.2em}
\label{pt}
\end{center}
\end{figure*}

\section{Performance Evaluation}\label{sec:analysis}
Here, we present the analysis on the energy efficiency of the cooperative link selection approach (Adaptive Scheme) under a variety of scenarios. The adaptive scheme was implemented for a multiple-source, multiple-relay hybrid LAP-terrestrial communication system in order to evaluate the energy efficiency.

\subsection{Simulation Environment Description}\label{simulation_settings}

In order to implement the hybrid LAP-Terrestrial source and multiple-relay network model, we used the simulator Omnet++\cite{OMNET} with the INETMANET framework. In our model, we implement wireless nodes with a two-layer hierarchical protocol architecture with an application layer to generate traffic and run the algorithm and a PHY layer having two physical interfaces. A further description of the simulation model implemented as separate modules is given below:

\begin{itemize}

\item[(i)] {\bf The Application Layer module:} This module simulates the Application Layer together with the adaptive scheme selection algorithm. The {\it adaptive scheme} is implemented using Alg.~\ref{alg:link} for the source node $S_j$ and Alg.~\ref{alg:link2} for the relay nodes $R_i$. The context-aware capabilities of the source and relay nodes are used in order to collect information needed for implementing the adaptive scheme approach.

\item[(ii)] {\bf The Physical Layer module:} It is composed of two interfaces to communicate with terrestrial and LAP node:

\begin{itemize}
\item[1)] The first physical interface is used to implement the terrestrial-terrestrial link (T-T) on ground using BPSK modulation in a Rayleigh fading channel with a pathloss coefficient as described in Sec.~\ref{sec:models}.
\item[2)] The second physical interface is used to implement the terrestrial-LAP (T-L) link using DPSK modulation in a Ricean channel with a pathloss coefficient as described in Sec.~\ref{sec:models}.
\end{itemize}

\item[(iii)] {\bf The Blackboard module:} This module emulates the context-aware capabilities of the nodes. It allows the nodes to share their information with the other nodes in the network. Therefore, the $Blackboard$ module contains information about the position of the nodes, the channel conditions of different links and others.
\end{itemize}

\subsection{Simulation Scenario and Results for Single-Source Multiple-Relay Hybrid LAP-terrestrial Communications}\label{simulation_settings}

\begin{table}[t]
\renewcommand{\arraystretch}{1}
\small
\centering
\caption{Parameters for channels characterization.}
\label{tbl:model}
\begin{tabular}{|l|c|c|c|c|c|}\hline
\textbf{Link}            &\textbf{Model}  & \textbf{d}    & \textbf{K}  & \textbf{$\alpha$} & \textbf{h}    \\
                         &                & \textbf{[m]}  & \textbf{[dB]}           &    &   \textbf{Dist.} \\
\hline
\hline
 $S_1 \rightarrow LAP$  & Ricean          &     4030.14    &  5         & 2.1 & Ricean \\
 $S_1 \rightarrow R_1$  & Rayleigh        &     1000.00    & -         & 2.0 & Rayleigh \\
 $R_1 \rightarrow LAP$  & Ricean          &     4030.14    &  6         & 2.0 & Ricean \\
 $S_1 \rightarrow R_2$  & Rayleigh        &     860.233    & -         & 2.1 & Rayleigh\\
 $R_2 \rightarrow LAP$  & Ricean          &     3504.71    &  8         & 2.0 & Ricean \\
 $S_1 \rightarrow R_3$  & Rayleigh        &     538.516    & -         & 2.3 & Rayleigh \\
 $R_3 \rightarrow LAP$  & Ricean          &     3799       &  10        & 2.1 & Ricean \\
\hline
\end{tabular}
\end{table}

\begin{figure*}[t]
\begin{center}
\subfloat[Cumulative Distribution Function (CDF)]{\includegraphics[width=0.50\textwidth]{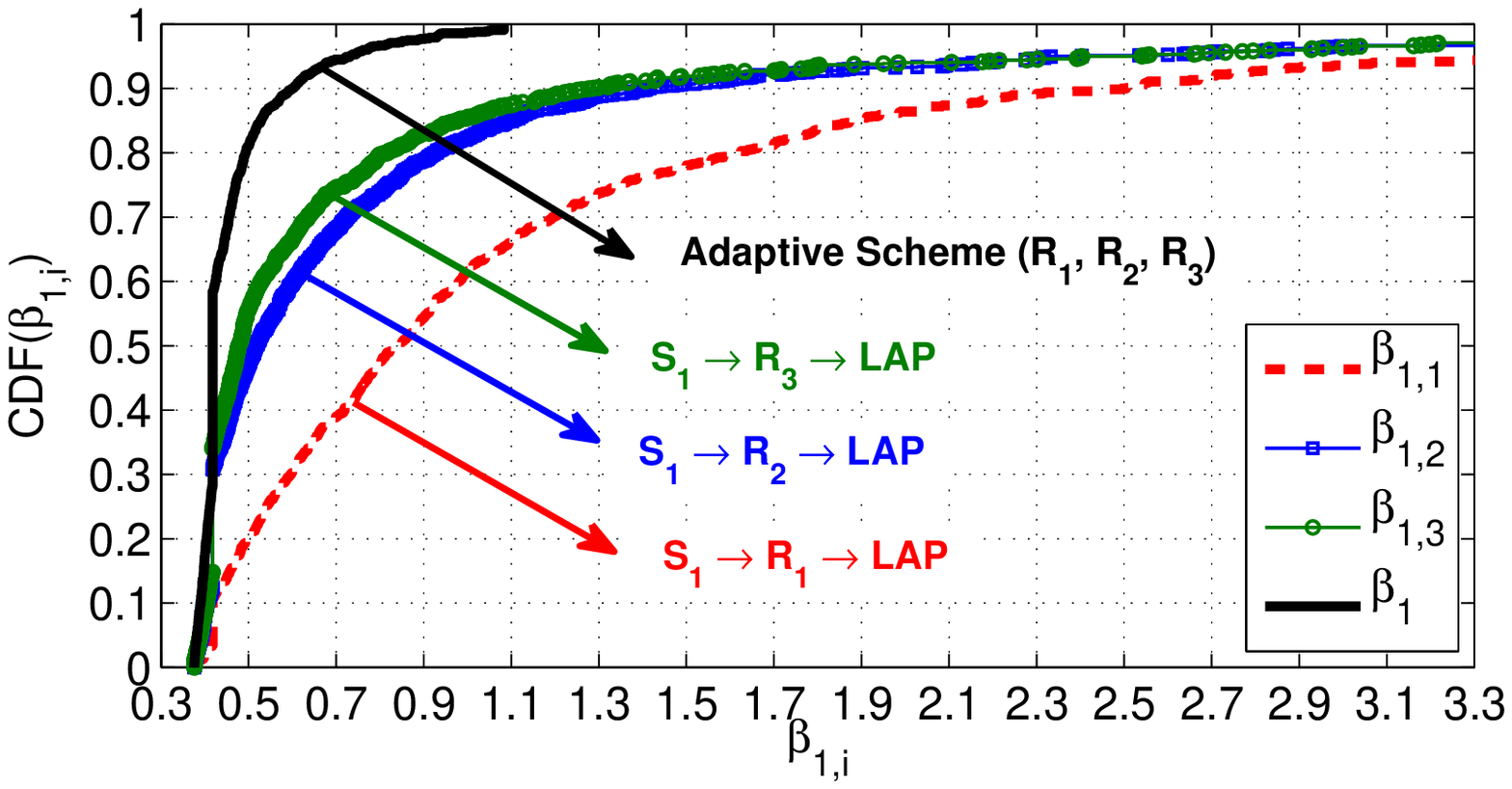}}
\subfloat[Probability Distribution Function (PDF)]{\includegraphics[width=0.50\textwidth]{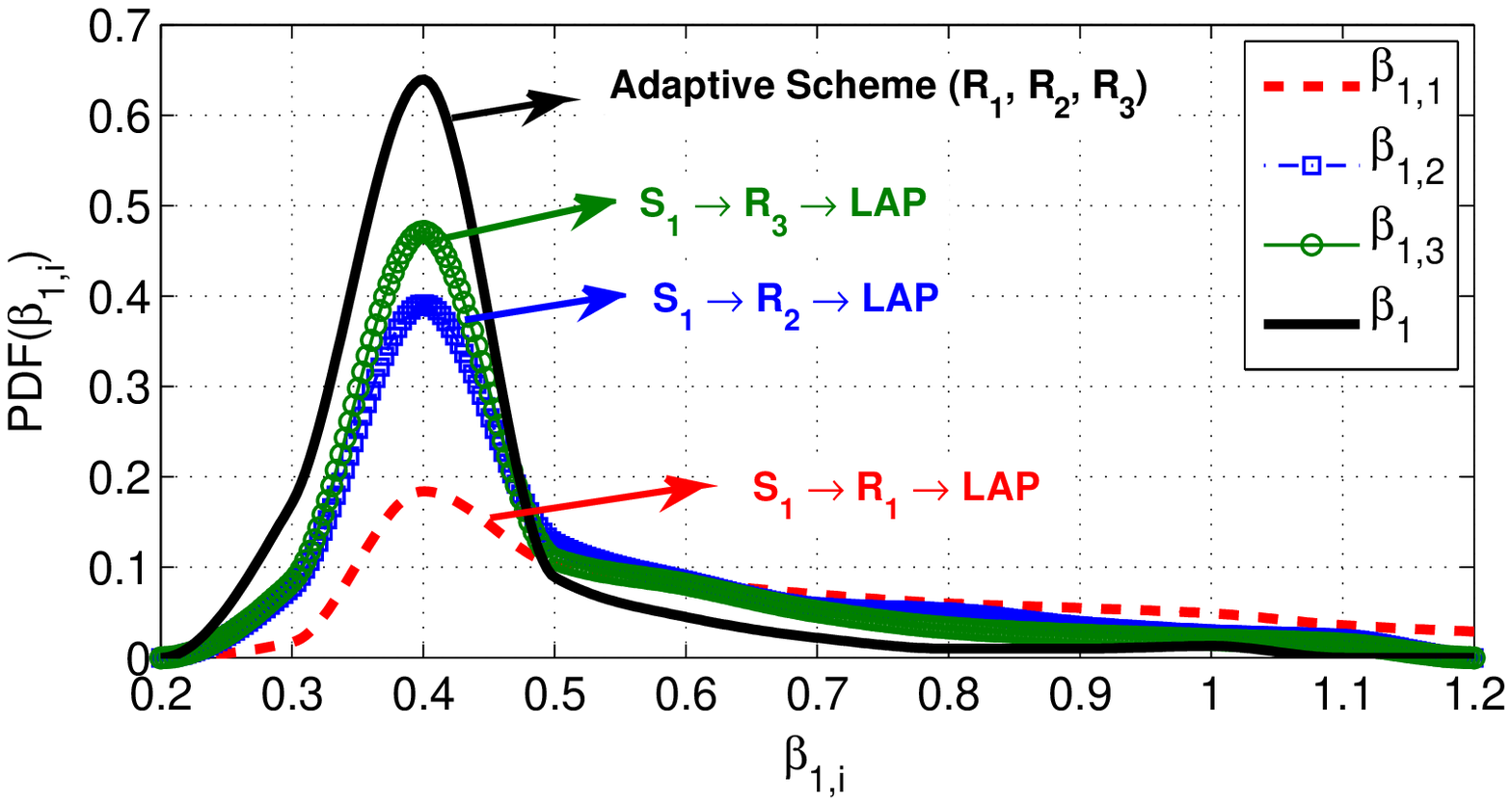}}
\caption{Comparing the energy efficiency between the direct and adaptive scheme by observing the CDF and PDF of $\beta_{j,i}$ for single-source multiple-relay hybrid LAP-terrestrial communications.}\vspace{-1.3em}
\label{Beta}
\end{center}
\end{figure*}

We considered a realistic disaster scenario with a perimeter block surface of 1~$km^2$, which represents $\approx$80 urban blocks. It is assumed that a natural disaster (e.g. a tsunami or an earthquake) destroyed the terrestrial communication network and electrical network distribution infrastructure. The hybrid aerial-terrestrial network is deployed for the public safety personnel and first responders in order to coordinate rescue and first-aid services for the survivors. The LAP is deployed in the center of the disaster area. The simulation scenario for the hybrid LAP-Terrestrial system with single-source and multiple-relay nodes is depicted in Fig.~\ref{fig5}. The scenario considers a total of $5$ nodes simulating one source node $(S_j)$, three relay nodes $(R_i)$, and one destination $(LAP)$ node. The parameters characterizing the wireless channels in the network are summarized in Table~\ref{tbl:model}. The chosen pathloss exponents are suitable for airborne communication networks (see \cite{alpha} for details). Some of the other parameters used in the simulations for all the nodes are: maximum transmission power equal to $2$~W, data-rate equal to $6$~Mbps, carrier frequencies for both terrestrial-terrestrial and terrestrial-LAP links equal to $3.5$~GHz, transmit and receive antenna gains equal to $3$~dB, the thermal noise power equal to $-125$~dBm, receiver sensitivity equal to $-100$~dBm. The power required for reception is set equal to $100$~mW for the terrestrial nodes and $200$~mW~\cite{GREENCOM1}, since energy spend in reception is about 50\% - 80\% less than energy spend in transmission. In order to inject traffic in the network, the Application Layer of the source node $S_j$ sends 1000 UDP packets of $1000$ bytes each for a total time of $1000$ seconds, each transmitted every second. The channel coherence time is also set to 1 sec (i.e. $T_{coh}=1$.

Fig.~\ref{pt} presents the total transmit power in time calculated by the direct and cooperative links prior to applying the adaptive scheme for single-source multiple-relay hybrid LAP-terrestrial communications. Here, the total transmit power level for the direct transmission was kept as a constant value of $P^1_t(1)=1000$~mW, and the transmit power levels for the source and the relay nodes ($P^{j,i}_t(2)+P^{j,i}_t(3)$) for cooperative transmission were iteratively computed for every packet based on the channel conditions (note that we assume slow flat fading channels, for which the magnitude of the channel can be considered constant over a packet duration, but changes over time). We observe that the cooperative link shows energy-efficient transmissions compared to the direct link as a consequence of better channel conditions in the cooperative link compared to what is observed in the direct link. Thus, we implement the adaptive scheme selection algorithm selecting between the direct and all possible cooperative links to choose the most energy-efficient link to reach the destination.

We study the energy efficiency of the adaptive scheme selection algorithm by computing ($\beta_{1}$) and compare it with $\beta_{j,i}$ for all the cooperative links without the adaptive scheme (i.e. comparing with $\beta_{1,1}$,$\beta_{1,2}$ and $\beta_{1,3}$). The corresponding cumulative distribution and probability density functions for $\beta$ are shown in Fig.~\ref{Beta}.a and Fig.~\ref{Beta}.b, respectively. In Fig.~\ref{Beta}.a, we observe that the energy efficiency factors $\beta_{1,1}$, $\beta_{1,2}$ and $\beta_{1,3}$ are below one for 65\%, 85\%, 88\% of the time respectively, indicating that the cooperative links are energy-efficient compared to the direct link. On the other hand, we observe that the energy efficiency factor $\beta_{1}$ for the adaptive selection method is under one for most of the time. From Fig.~\ref{Beta}.b, we can observe that there is a higher probability that $\beta_{1,i}$ values are concentrated around less than one (more or less $0.3-0.5$) for the assumed scenario. Therefore, we can intuitively argue that the cooperative link selection-based transmission gives better energy efficiency.

\begin{figure}
\begin{center}
\includegraphics[width=0.50\textwidth]{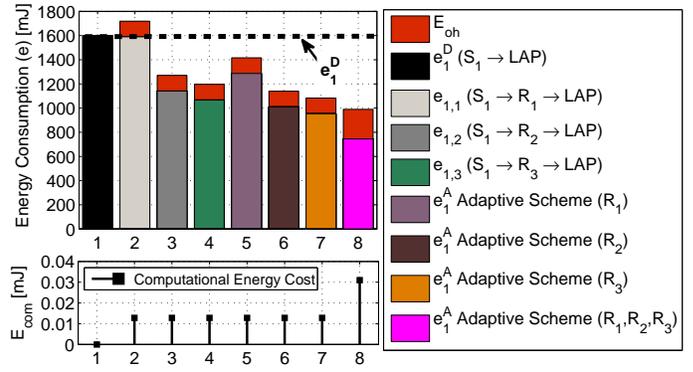}
\caption{Energy consumption of the network for the direct link, cooperative links and adaptive scheme for single-source multiple-relay hybrid LAP-terrestrial communications. Context-aware packet reporting period $\tau_R =$ 200 ms.}
\label{energy}
\end{center}
\end{figure}

Based on the expressions in Section~\ref{energy_comp}, we compute the total energy consumption for the direct link $e_j^D$, the relay links $e_{j,i}$ and the adaptive scheme $e^A_j$ for the single-source multiple-relay hybrid LAP-terrestrial communication. Fig.~\ref{energy} depicts the energy consumption results for the direct and the adaptive schemes. In the figure, we also present the energy consumption values for adaptation between the direct link and with only one relay node $R_i$ as well as for a fully adaptive system where the adaptation is performed between the direct link and all the relay links. It is important to note that the context-aware packet energy cost is calculated using the equations presented in Appendix~\ref{sec:appendix}.1 and the reporting period of the context-aware packets is $\tau_R =$ 200 ms. In Fig.~\ref{energy}, we observed the following:

\begin{itemize}
\item[1)] The energy consumption for the context-aware packets has huge impact on the energy consumed using the adaptive scheme;
\item[2)] The energy consumption for the relay link $e_{1,2}$ and $e_{1,3}$ is lower than the energy consumption for the direct link, while the energy consumption for relay link $e_{1,1}$ is higher than the energy consumption efor the direct link;
\item[3)] The energy consumption for the direct link $e_1^D$ is the higher than the energy consumption for the adaptive scheme $e^A_{1}(R_1)$, $e^A_{1}(R_2)$, $e^A_{1}(R_3)$ and $e^A_{1}(R_1,R_2,R_3)$;
\item[4)] The energy consumption for the adaptive scheme $e^A_{1}(R_1)$, $e^A_{1}(R_2)$ and $e^A_{1}(R_3)$, using only one relay, improves the energy consumption for each relay link;
\item[5)] The energy consumption for the fully adaptive scheme $e^A_{1}(R_1,R_2,R_3)$ using all the relays nodes $R_i$ is the lowest in the network;
\end{itemize}

Note that the above results are based on the assumed simulation values. In particular, for the wireless channels there can also be situations where the direct link is less energy-efficient than some or all of the relay links, in any case the adaptive scheme selection technique will provide energy-efficient transmissions by utilizing the channel variations in time for a given bit error probability. Finally, the Fig.~\ref{energy} also shows the total computational energy cost ($E_{com}$) of the adaptive scheme, which is computed using the equations presented in Appendix~\ref{sec:appendix}.2. As it can be seen, the total $E_{com}$ represent less than 1\% of the total energy consumption of the network the for all the scenarios.
\begin{figure}
\begin{center}
\includegraphics[width=0.50\textwidth, height=45mm]{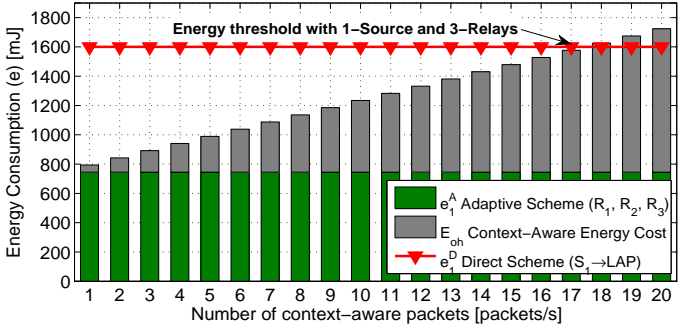}
\caption{Total energy consumption of the network for the direct and adaptive scheme versus the number of context-aware packet per second (1-source, 3-relays).}
\label{energyoh}
\end{center}
\end{figure}

Fig.~\ref{energyoh} depicts the total energy consumption of the network for the direct and adaptive scheme in comparison with the number of transmitted context-aware packet per second. The adaptive scheme energy cost is calculated using the single-source that is transmitting 100 packets/s using 3-relays. Here, the context-aware reporting packets energy cost is calculated using the equations presented in Appendix~\ref{sec:appendix}.1 and a context-aware packet reporting time is increasing from 50 up to 1000 ms. The {\it energy threshold} is indicated in the figure, which is the point where the direct link is more energy efficient than the adaptive scheme, for the simulation settings used in this paper the energy threshold point is achieved when the context-aware packet reporting time is less than $\approx$ 60 ms.
\begin{figure}
\begin{center}
\includegraphics[width=0.45\textwidth, height=45mm]{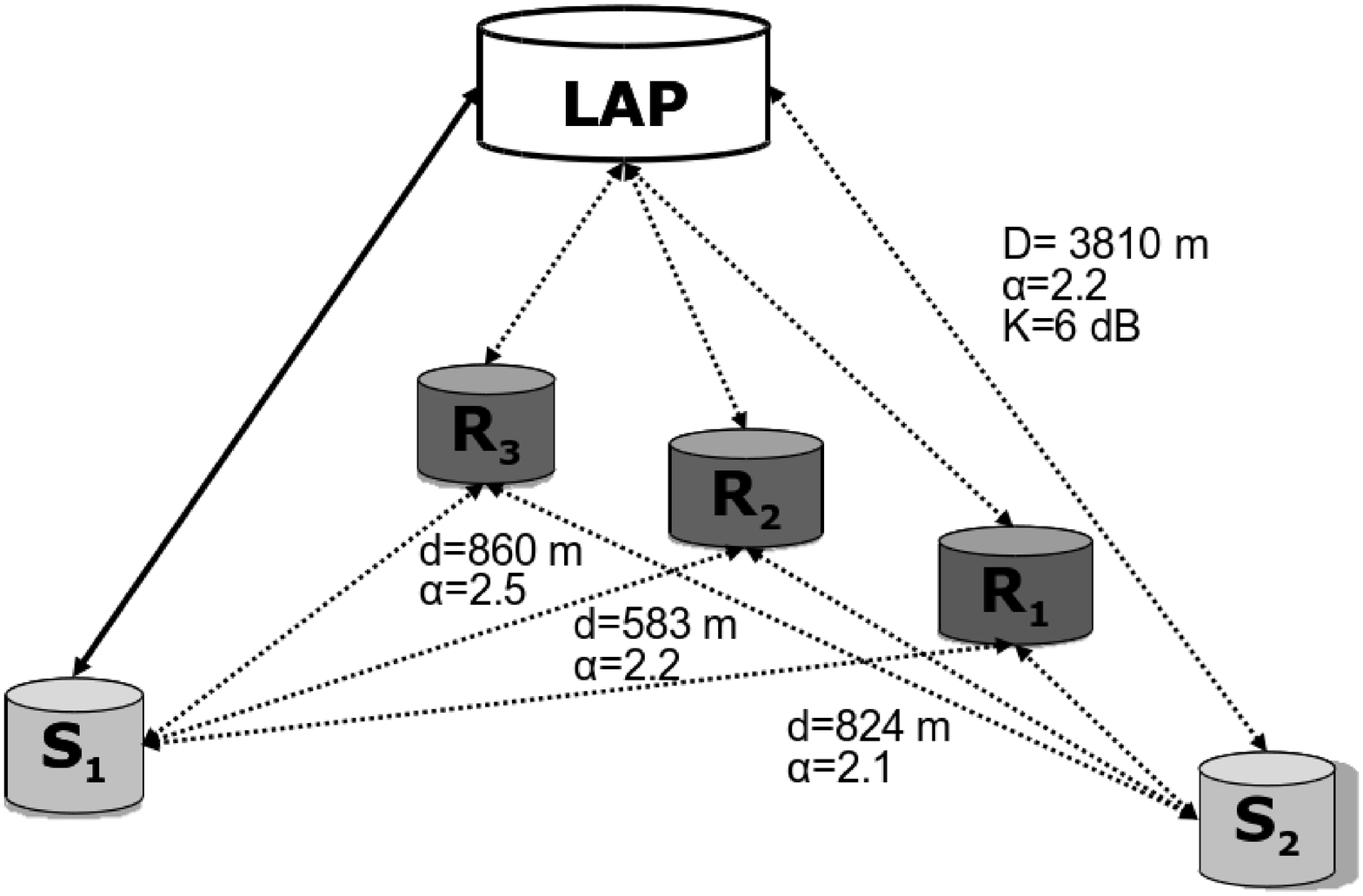}
\caption{Scenario for multiple-source multiple-relay hybrid LAP-terrestrial communications.}
\label{fig6}
\end{center}
\end{figure}

\begin{figure}
\begin{center}
\subfloat[Total $e$ at network level]{\includegraphics[width=0.25\textwidth, height=48mm]{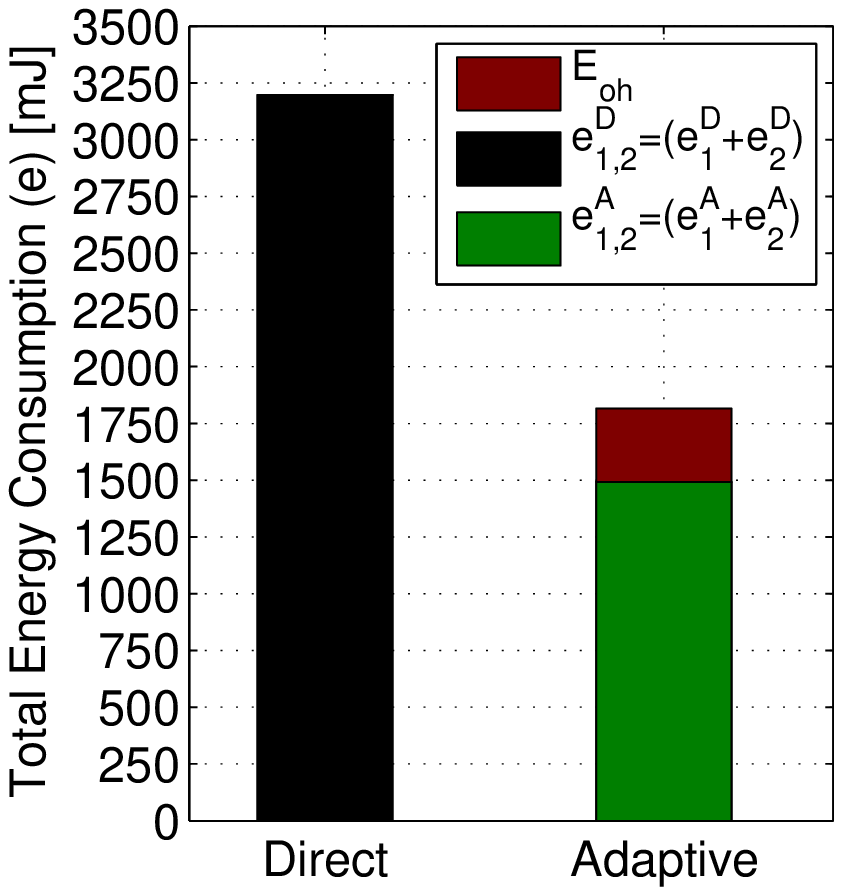}}
\subfloat[Average $e$ at source level]{\includegraphics[width=0.25\textwidth, height=48mm]{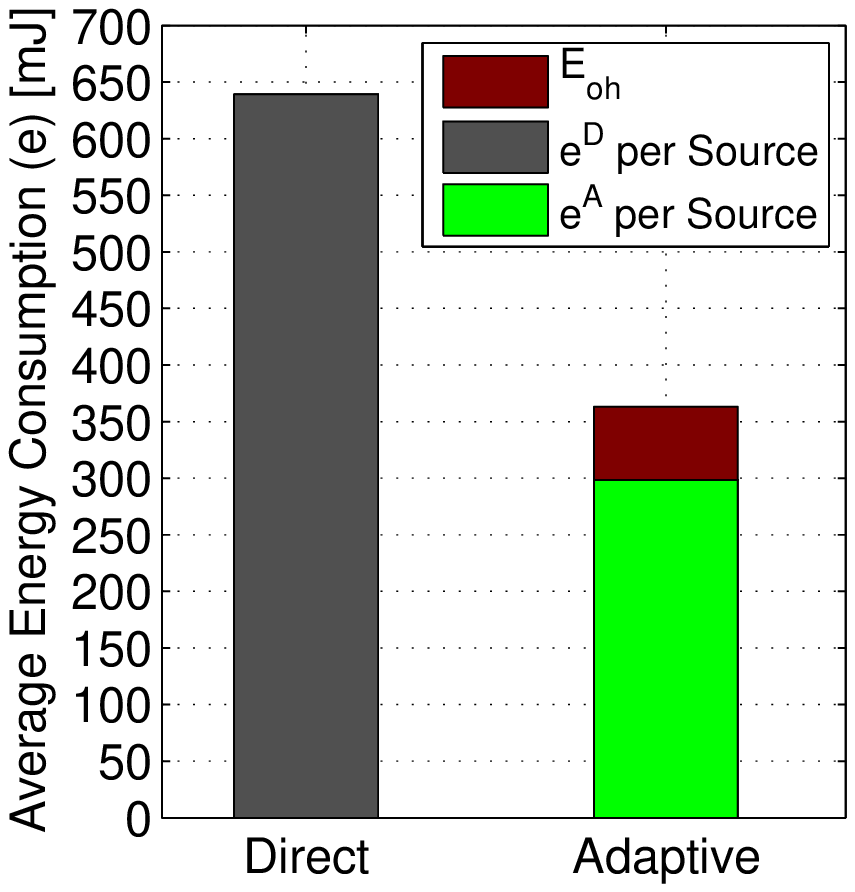}}
\caption{Energy consumption for the direct link, cooperative links and adaptive scheme for multiple-source multiple-relay hybrid LAP-terrestrial communications.}
\label{energy2}
\end{center}
\end{figure}

\subsection{Simulation Scenario and Results for Multiple-Source Multiple-Relay Hybrid LAP-terrestrial Communications}\label{simulation_settings}
In order to evaluate the scenario for the multiple-source and multiple-relay nodes in the hybrid aerial-terrestrial network, we keep the same settings for the simulations from the previous section. For the multiple source scenario we initially consider only two source nodes with three relay nodes and then extend our analysis to ten source nodes and six relay nodes as discussed subsequently.

\begin{table}[t]
\renewcommand{\arraystretch}{1}
\small
\centering
\caption{Parameters for channels characterization.}
\label{tbl:model2}
\begin{tabular}{|l|c|c|c|c|c|}\hline
\textbf{Link}            &\textbf{Model}  & \textbf{d}    & \textbf{K}  & \textbf{$\alpha$} & \textbf{h}    \\
                         &                & \textbf{[m]}  & \textbf{[dB]}           &    &  \textbf{Dist.}  \\
\hline
\hline
 $S_2 \rightarrow LAP$  & Ricean          &     3810    &  6         & 2.2 & Ricean  \\
 $S_2 \rightarrow R_1$  & Rayleigh        &     824    & -         & 2.1 & Rayleigh  \\
 $S_2 \rightarrow R_2$  & Rayleigh        &     583    & -         & 2.2 & Rayleigh  \\
 $S_2 \rightarrow R_3$  & Rayleigh        &     860    & -         & 2.5 & Rayleigh \\

\hline
\end{tabular}
\end{table}

\begin{figure}
\begin{center}
\includegraphics[width=0.50\textwidth, height=45mm]{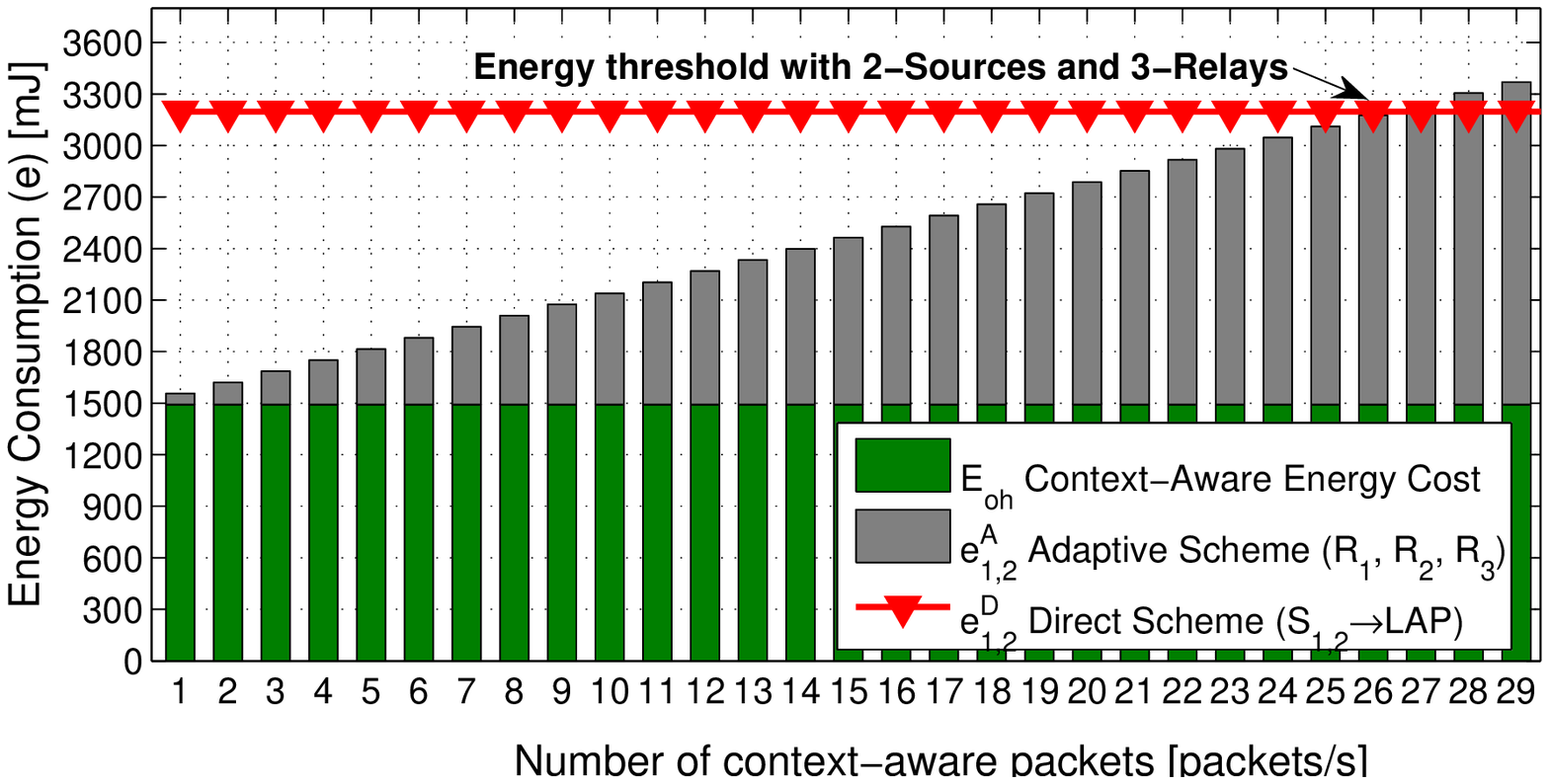}
\caption{Total energy consumption of the network for the direct and adaptive scheme versus the number of context-aware packet per second (2-source, 3-relays).}
\label{energyoh2}
\end{center}
\end{figure}

\begin{table}[t]
\renewcommand{\arraystretch}{1}
\small
\centering
\caption{Parameters for channels characterization.}
\label{tbl:model3}
\begin{tabular}{|l|c|c|c|c|}\hline
\textbf{Link}            &\textbf{Channel Model}    & \textbf{K [dB]}  & \textbf{$\alpha$}   \\
\hline
\hline
Uplink Channel         & Ricean           &  4 - 10 & 2.0 - 2.2     \\
Terrestrial  Channel  & Rayleigh       & -              & 2.1 - 2.8 \\
\hline
\end{tabular}
\end{table}

Fig.~\ref{fig6} depicts the simulation scenario for the two source node and 3 relay nodes scenario. The parameters for characterizing the additional channels involving the new node $S_2$ in the network are summarized in Table~\ref{tbl:model2}. The total and average energy consumption for the two direct links given by $e_{1,2}^D = e_1^D +e_2^D$, and the adaptive scheme $e_{1,2}^A = e^A_1 + e^A_2$ for the multiple-source multiple-relay network were calculated from the simulations and the corresponding results are shown in Fig.~\ref{energy2}. As expected, we observe from Fig.~\ref{energy2}.a that the total energy consumption for the direct links $e_j^D$ is much higher than the adaptive scheme $e_j^A$. The same behavior is observed in Fig.~\ref{energy2}.b where the average energy consumption per node is shown. Note that the context-aware reporting period is $\tau_R=$ 200 ms.

Fig.~\ref{energyoh2} depicts the total energy consumption of the network for the direct and adaptive scheme in comparison with the number of transmitted context-aware packet per second. The adaptive scheme energy cost is calculated using 2-source transmitting 100 packets/s using 3-relays. The {\it energy threshold} is indicated in the figure, the energy threshold point is achieved when the context-aware packet reporting time is less than $\approx$ 40 ms. It is important to note that when the number of sources in the network is increased the context-aware packet energy cost is reduced because the introduced overhead is sharing between the multiple-sources.

Finally, in order to evaluate a more complex scenario for the multiple-source and multiple-relay nodes in the hybrid aerial-terrestrial network, we keep the same settings for the simulations from the previous section for the transmission and reception power levels and the antennas gains. Then, the network is configured for a total of $17$ nodes with $10$ source nodes $(S_j)$, $6$ relay nodes $(R_i)$ and $1$ destination $(LAP)$ node. The parameters for characterizing the channels and nodes position are summarized in Table~\ref{tbl:model3}. The channel parameters and the position of the terrestrial nodes were all assigned randomly over a prescribed range rather than pre-assigning them. All the terrestrial nodes are randomly placed within the $[1000,1000]m^2$ ground while the LAP position is fixed at at the center of the ground with an altitude of $2$km. The pathloss exponents for the uplinks (terrestrial to LAP) are randomly assigned between $2.0$ and $2.2$ (uniformly distributed), the pathloss exponents for the terrestrial links are randomly assigned between $2.1$ and $2.8$ (uniformly distributed), and all $K$ (rice factor) values for the uplink are randomly assigned between $4$~dB and $10$~dB. Note that the range of values selected for the pathloss exponents and the rice factors are based on practical conditions \cite{Rapp},\cite{MastersThesis}, \cite{alpha}, such random selections of parameters are done to avoid biased channel gain assignments which would give us a fair analysis.

\begin{figure}
\begin{center}
\subfloat[$e$ per Source]{\includegraphics[width=0.50\textwidth, height=45mm]{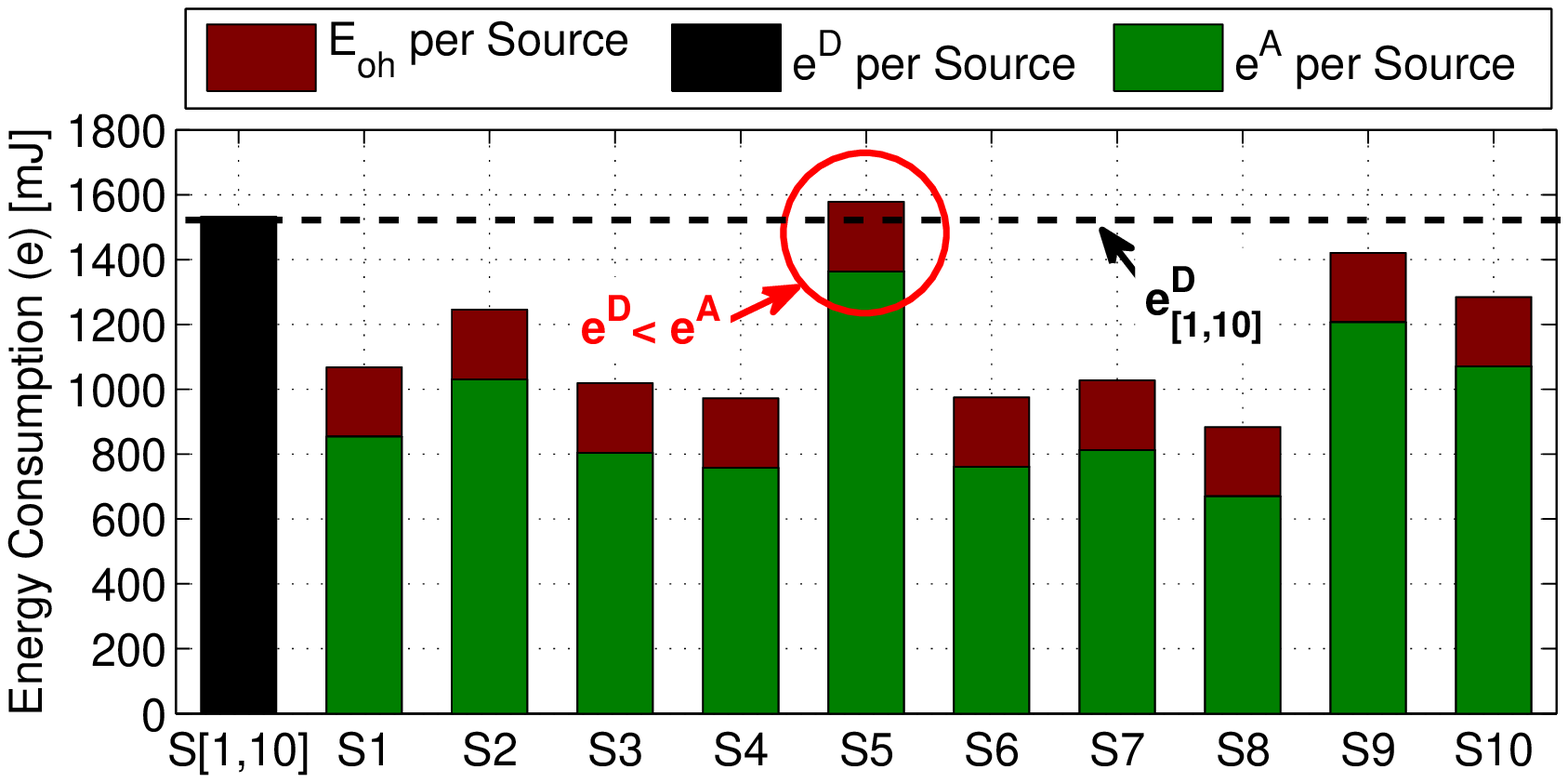}}\\ \vspace{-0.8em}
\subfloat[Total $e$ at network level]{\includegraphics[width=0.25\textwidth, height=40mm]{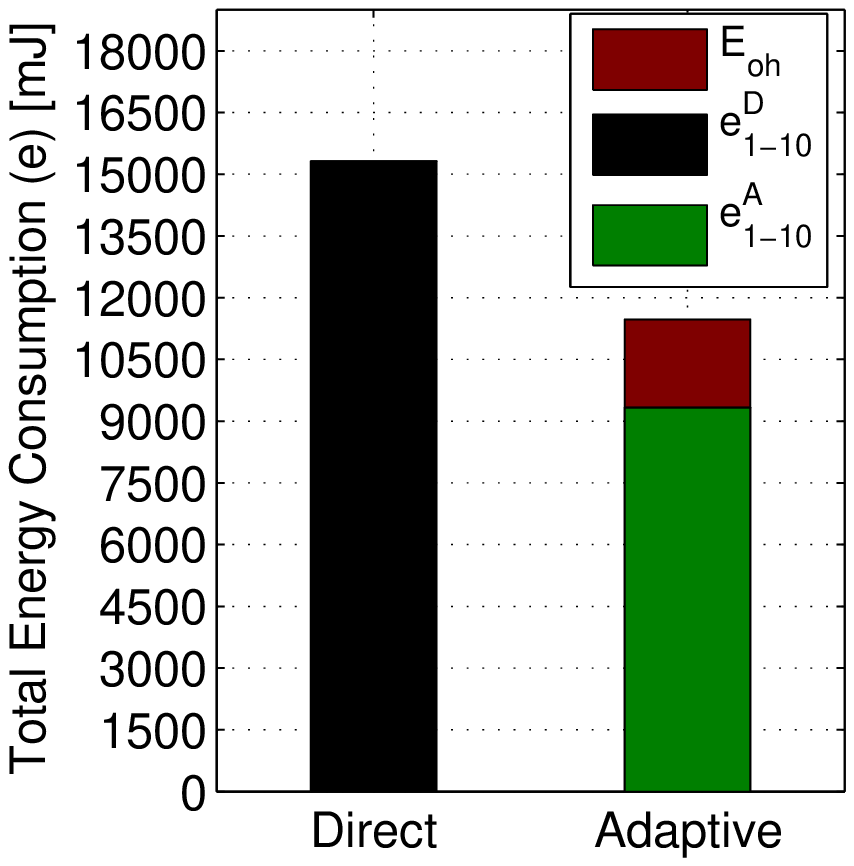}}
\subfloat[Average $e$ at source level]{\includegraphics[width=0.25\textwidth, height=40mm]{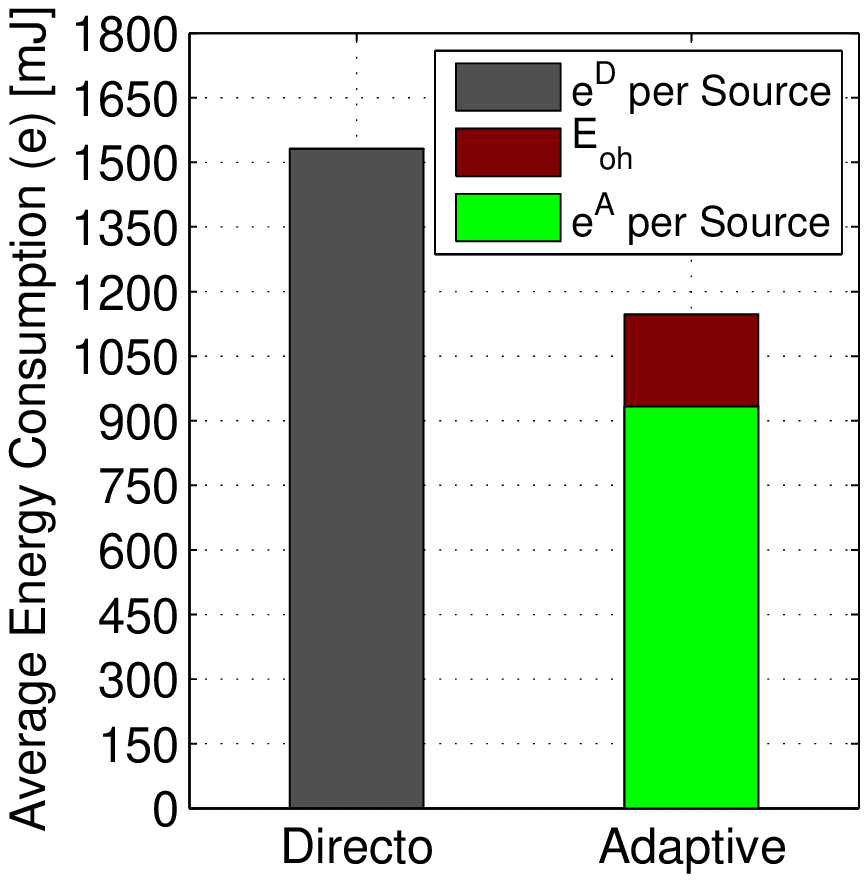}}
\caption{Total energy consumption per source for the direct link and adaptive scheme for multiple-source multiple-relay hybrid LAP-terrestrial communications. Context-aware packet reporting period $\tau_R =$ 200 ms}
\label{energysource}
\end{center}
\end{figure}

Based on the expressions in Section~\ref{energy_comp}, we compute the total energy consumption for the direct link $e_j^D$ and the adaptive scheme $e^A_j$ per sources. Fig.~\ref{energysource} depicts the energy consumption results for the direct and adaptive schemes for a context-aware packet reporting period of 200 ms. In Fig.~\ref{energysource}.a, we observed that the energy consumption for the direct link $e_{1-10}^D$ is the higher than the energy consumption for the adaptive scheme $e^A_{1-10}(R_1-R_6)$ in most of the cases. However, it is important to note that over the $10$ sources only in the case of source number $5$ ($S_5$), the energy consumed by direct link is better than the energy consumed by adaptive scheme. In Fig.~\ref{energysource}.b-c, the total and energy consumption for the direct link $e_j^D$ and the adaptive scheme $e^A_j$ are also shown in order to remark the energy saving achieved using the adaptive scheme.

\begin{figure}
\begin{center}
\includegraphics[width=0.50\textwidth, height=45mm]{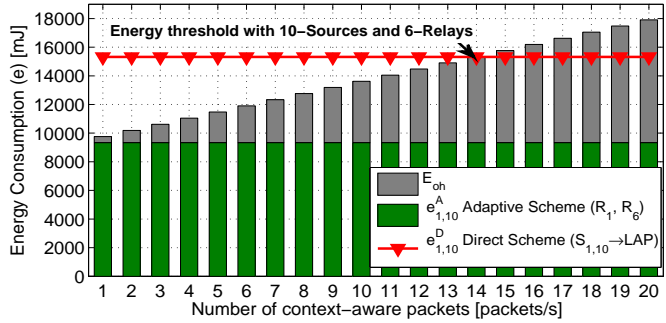}
\caption{Total energy consumption of the network for the direct and adaptive scheme versus the number of context-aware packet per second (6-source, 3-relays).}
\label{energyoh3}
\end{center}
\end{figure}

Fig.~\ref{energyoh3} depicts the total energy consumption of the network for the direct and adaptive scheme in comparison with the number of context-aware packet per second. The adaptive scheme energy cost is calculated using 10-source transmitting 100 packets/s using 6-relays. The {\it energy threshold} is indicated in the figure, the energy threshold point is achieved when the context-aware packet reporting time is less than $\approx$ 70 ms.

\subsection{Simulation Scenario and Results for Cooperation Delay in Multiple-Source and Multiple-Relay Network.}\label{delays}

In this section, the corresponding analysis for the delay is presented by considering a single service queue model for all the (relaying) nodes. The proposed relay selection algorithm is optimized for energy efficiency but not for delay. The delay analysis here is essential since the relay nodes $R_i$ can also act as source nodes to transmit their own messages, in which case a particular node will have a higher priority for its own traffic and a lower priority for the relay traffic received from other neighboring nodes. The relay nodes will then delay the relay messages from other source nodes by storing them in a queue $Q$. It is important to remark that the $Q$ delay considered here does not include the access or retransmission delays in the MAC layer but only considers the queuing delay with some random finite processing/servicing time of the packets or both classes of traffic (i.e. high priority class for local traffic and lower priority class for relay traffic). The delay introduced due to the MAC and Network Layers are not considered here as our primary motivation is to study the degraded delay performance due to the relaying process in the hybrid aerial-terrestrial system.

\begin{figure}
\begin{center}
\subfloat[2-sources and 3-relays]{\includegraphics[width=0.45\textwidth, height=40mm]{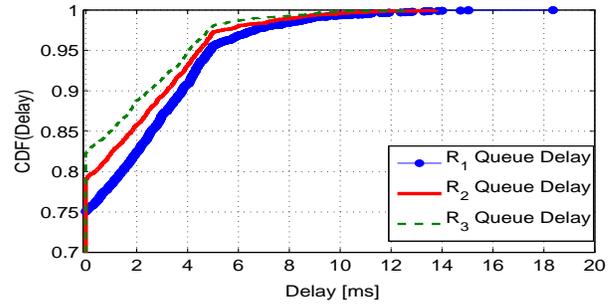}}\\
\subfloat[10-sources and 6-relays]{\includegraphics[width=0.45\textwidth, height=40mm]{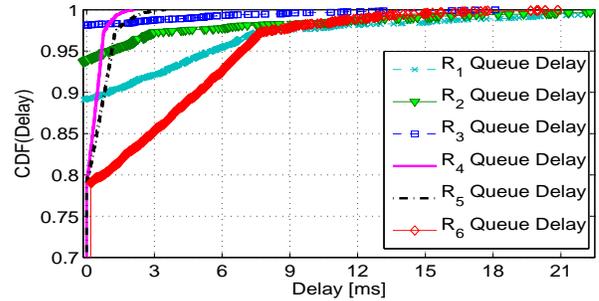}}
\caption{Cumulative Distribution Function of the delay introduced for the queue created for the cCooperation in Multiple-Source and Multiple-Relay Network.}
\label{delay}
\end{center}
\end{figure}

For the delay analysis here the source traffic generated at every node (including a node that acts as a relay) is modeled by an exponential inter-arrival traffic with a mean inter-arrival time of $20$~ms and a queue service time of $5$~ms to serve the higher priority traffic at any node. The relay traffic however would have a different (scenario-specific) traffic model. The relay traffic received by a relay node in this case depends on the selection of that particular node as a relay by other nodes and therefore the most energy-efficient relay node here will have a higher arrival rate for the relay traffic and consequently the least energy-efficient relay node will have a lower arrival late. In this sense one would expect a higher delay at the most energy-efficient relay node. The simulation results shown in Table~\ref{tbl:results} and Fig.~\ref{delay} verify this.

Table~\ref{tbl:results} basically shows (i) {\bf QueueingTime:mean} which is the average delay introduced by the queue in each message and (ii) {\bf QueueingTime:max} which is the maximum delay introduced by the queue to the message while Fig.~\ref{delay} shows the cumulative distribution function (CDF) of the delays at every node. The relay node with the higher delay corresponds to the most energy-efficient path to the destination node (in this case relay node $R_3$). Therefore, we clearly see a trade-off between the energy-efficient cooperative scheme and delay in the strategy that we have proposed. The delay however is not significant enough according to the simulation results to affect the relaying traffic and based on the results, a real-time traffic such as voice could be easily supported. But on the other hand, this of course depends on the arrival rate and the relay selection operation based on channel quality factors.

\begin{table}
\renewcommand{\arraystretch}{1}
\centering
\small
\caption{Simulation Results for the delay analysis.}
\label{tbl:results}
\begin{tabular}{|l|l|c|c|}
\hline
\textbf{Node}      &\textbf{Queueing}           &  \textbf{Queueing}\\
                   & \textbf{Time (Mean)}       & \textbf{Time (Max)} \\
\hline
\hline
Relay$_1$          & 5.4809$\times 10^{-04}$ s  & 0.0128 s \\	
\hline
Relay$_2$         & 6.9067$\times 10^{-04}$ s   & 0.0138 s    	  \\	
\hline
Relay$_3$        & 8.8219$\times 10^{-04}$ s    & 0.0184 s  	  \\	
\hline
\end{tabular}
\end{table}

\section{Conclusions and Future Work}\label{sec:conclusion}
In this paper, we discussed a new integrated and hybrid aerial-terrestrial communication architecture for emergency communications. We proposed a real-time adaptive transmission scheme (adaptive scheme) which dynamically selects the best link based on the channel conditions for enabling energy-efficient communications in hybrid aerial-terrrestrial networks. We showed that the cooperation between mobile terrestrial terminals on the ground improves the energy efficiency in the uplink depending on the temporal behavior of the terrestrial and aerial uplink channels; at the same time the delay associated with relaying traffic also increases at the most energy-efficient relay node. The simulation results confirmed that the adaptive transmission scheme shows efficient and reliable performance with respect to the energy consumption when compared to cooperative transmission and direct transmission techniques. We also investigate the impact of the introduced overhead and computational operations on the energy consumption of the network. We have observed that there is a threshold between the energy saved with the adaptive scheme and the energy consumed for implementing the adaptive scheme, which is determinate by the context-aware reporting packet time. 

We are currently extending our adaptive transmission and energy-efficient link selection approach to analyze cooperative strategies in network topologies with multiple LAPs. Besides, we are evaluating possible adaptations for the influence of battery capacities of terrestrial terminals on relay selection in multi-LAP aerial-terrestrial systems for efficient transmissions considering game-theoretic approaches.

\section*{Appendix - A}\label{sec:appendix}
The two types of overhead energy costs $E_{oh}$ and $E_{com}$ described in Section-V-C are described in detail below.\\

\emph{A.1. Estimated Context-Aware Energy Cost ($E_{oh}$)}\\
There are several algorithms for exchanging the context-aware information between the LAP, source and relay nodes as we previously explained. Here, we introduce a general energy model for estimating the average context-aware energy cost assuming that broadcast packets are used. The total context-aware energy cost is given by,
\begin{equation}\label{eq151}
E_{oh} = E_{A-oh}+E_{T-oh}
\end{equation}
\noindent where, $E_{A-oh}$ is the energy cost of aerial information dissemination and $E_{T-oh}$ is the energy cost for the terrestrial information dissemination. It is assumed that all the a broadcast report is received by all the nodes in the network, therefore $E_{A-oh}$ and $E_{T-oh}$ are respectively given by
\begin{equation}\label{eq152}
E_{A-oh}=(P^{LAP}_{Tx}+M \times P_{Rx})\times T_{ca}
\end{equation}
\begin{equation}\label{eq153}
E_{T-oh}=M\times(P_{Tx}+(M-1)\times P_{Rx})\times T_{ca}
\end{equation}
\noindent where, $T_{ca}$ is the total time taken for context-aware data transmission in a period of $T$, and as describe in the previous sections $P^{LAP}_{Tx}$ is the transmit power of the LAP terminal, $P_{Tx}$ is the transmit power for any terrestrial node, $P_{Rx}$ is the required power to receive data at any terrestrial node, and finally $M$ is the total number of terrestrial nodes in the system.\\

Finally, the energy cost for the context-aware packets is calculated using the information needed for computing the adaptive power control algorithm. In the reporting packets, the following information is included:

\begin{itemize}
\item The header, which includes the address of the source that broadcasts the packet 32 bits,
\item The transmit EIRP (i.e. $P_tG_t$), $\approx$128 bits,
\item The location coordinates for GPS position, $\approx$128 bits,
\end{itemize}

\noindent Then, the total packet length is estimated in a total of $\approx 288$ bits. Note that the context-aware packets are transmitted with a constant transmission power level equal to 199.5~mW (or 23~dBm) for the terrestrial nodes and $1000$~mW for the aerial node with the basic bitrate equal to 2~Mb/s for both cases.

\emph{A.2. Estimated Computational Energy Cost ($E_{com}$)}\\
In order to estimate the computational energy cost, we consider the energy model proposed in~\cite{comp1}, given by,
\begin{equation}\label{eq100}
E_f = P_f\times T_f
\end{equation}
\noindent where, $E_f$ is the average energy consumed by the system while possessing a single CPU intensive workload\footnote{CPU intensive workload are the instructions of a computer program, which are the basic arithmetical, logical, and input/output operations of the system.} (CPU-W), $P_f$ is the power usage of a CPU-W as function of the processor clock rate $\Omega$, and $T_f$ is the time taken for the system to perform a single CPU-W which is given by ($1/\Omega$).

The described energy model in (\ref{eq100}) for numerical computation is also suitable for modeling the CPU energy consumption of smart-phones~\cite{comp2,comp3}. Therefore, it is applicable to the terrestrial terminals ($S$ or $R$), which compute and execute the power control algorithm. Therefore, $E_{com}$ is given by,
\begin{equation}\label{eq101}
E_{com} = \Psi \times E_f
\end{equation}
\noindent where, $\Psi$ is the number of CPU-W identified for the adaptive power control algorithm at the source and relay nodes for computing the optimal transmit power level. Therefore, we use (\ref{eq101}) for estimating the computational energy cost for the cooperative link.

Firstly, we concentrate on the energy spend on the calculation of the parameters needed for computing the adaptive power control algorithm. The distance between the source and relay is given by:
\begin{eqnarray}\label{eq99}
d_i=\arccos[\cos(90-Lat_{S}) \cos(90-Lat_{R}) +  \nonumber\\ \sin(90-Lat_{S}) \sin(90-Lat_{R}) \nonumber\\ \times \cos(Lon_{S}-Lon_{R}))]\times 6371,
\end{eqnarray}

\noindent where, [$Lat_{S}$,$Lon_{S}$] and  [$Lat_{R}$,$Lon_{R}$] are the GPS coordinates of the source and relay position respectively ($d_i$ in km). While the received power of the context-aware packet $P_{r}$ is given by:
\begin{eqnarray}\label{eq999}
P_{r} = EIRP-L + G_{r},
\end{eqnarray}

\noindent where, $EIRP$ is the effective isotropically radiated power, $L$ is the path loss and $G_{r}$ is the receiver antenna gain as described in Section-IV. Thus the computational energy cost for calculating i) the distance ($d_i$) between source and relay from the GPS coordinates and ii) the received power of the context-aware packet are given by,

\begin{itemize}
\item {\it $d_i$ computation}: 14 arithmetical operations,
\item {\it Received power computation}: 3 arithmetical operations,
\end{itemize}

\noindent Therefore, the number of CPU-W for the parameters calculation is $\Psi_1=17$. Then, based on Alg.~\ref{alg:link} and~\ref{alg:link2} (Section~\ref{sec:algorithms}), we estimate $\Psi_2$ for computing and executing the adaptive power control algorithm as described below.
\begin{itemize}
\item {\it For the source node}: 10 arithmetical operations per packet plus 30 arithmetical and 2 logical operations per each relay node detected by the source node,
\item {\it For the relay node}: 1 logical and 2 input/output operations per packet.
\end{itemize}

\noindent Therefore, the number of CPU-W for the adaptive power control algorithm is $\Psi_2=13+32 M$, and thus (\ref{eq101}) using $\Psi_1$ and $\Psi_2$ becomes,
\begin{equation}\label{eq103}
E_{com} = (30+32M)\times E_f
\end{equation}

\noindent where, $M$ is the number of relay candidates. Note that $E_f$ depends of several factors related to hardware and software specification. For example, $E_f$ can be calculated using $P_f$=150~mW and $\Omega=$528~MHz (see~\cite{comp1} for details), which are typical values for HTC Dream Android smartphone (CPU ARM 11 @ 528~MHz).

\section*{Acknowledgments}

The research leading to these results has received partial funding from the EC Seventh Framework Programme (FP7-2011-8) under the Grant Agreement FP7-ICT-318632, ABSOLUTE Project.

%






\begin{thebibliography}{50}

\bibitem{WPMC2011}
L. Reynaud, T. Rasheed and S. Kandeepan, \emph{An Integrated Aerial Telecommunications Network that Supports Emergency Traffic}, in Proc. of WPMC EMT Workshop, October 2011.

\bibitem{pe-wasun}
L. Reynaud and T. Rasheed, \emph{Deployable Aerial Communication Networks: Challenges for Futuristic Applications}, in Proc. of ACM PE-WASUN, Cyprus, October 2012.

\bibitem{Deaton}
J. D. Deaton, \emph{High Altitude Platforms for Disaster Recovery: Capabilities, Strategies, and Techniques for Emergency Telecommunications}, EURASIP Journal on Wireless Communications and Networking, vol. 2008, 8 pages, 2008.

\bibitem{HAP_Book1}
A. Aragon-Zavala, J.L. Cuevas-Ruiz and J.A. Delgado-Penin, \emph{High-Altitude Platforms for Wireless Communications}, Ed. John Wiley and Sons, Ltd., United Kingdom, 2008.

\bibitem{TAES1}
John Marini, Error Analysis for Relay Type Satellite-Aided Search and Rescue Systems, IEEE Transactions on aerospace Electronic Systems, Vol. AES-15, No. 5 Sep 1979.

\bibitem{TAES2}
Pengcheng Zhan, Kai Yu and  A. Lee Swindlehurst, Wireless Relay Communications with Unmanned Aerial Vehicles: Performance and Optimization, IEEE Transactions on aerospace Electronic Systems, Vol. 47, No. 3 July 2011

\bibitem{TAES3}
Andrea Giorgetti, Matteo Lucchi, Marco Chiani, Moe Z. Win, Throughput per Pass for Data Aggregation from a Wireless Sensor Network via a UAV, IEEE Transactions on aerospace Electronic Systems, Vol. 47, No. 4 October 2011.

\bibitem{google}
Google's news. Broadband Internet Access to the Developing World. Available at \url{http://ravenaerostar.com/about/project-loon-raven-aerostar-google}.

\bibitem{del_re}
S. Morosi, S. Jayousi and E. Del Re, \emph{Cooperative Strategies of Integrated Satellite/Terrestrial Systems for Emergencies}, in Proc. of ICST on Personal Satellite Services, Rome-Italy, Feb. 2010.

\bibitem{ABSOLUTE}
ABSOLUTE (Aerial Base Stations with Opportunistic Links for Unexpected and Temporary Events), EU FP7 Integrated Project. Available at \url{http://www.absolute-project.eu}.

\bibitem{FP7_EARTH}
EARTH (Energy Aware Radio and Network Technologies), EU FP7 ICT Project. Available at \url{http://www.ict-earth.eu}.

\bibitem{PSATS11}
S. Kandeepan, T. Rasheed and S. Reisenfeld, \emph{energy-efficient Cooperative HAP-Terrestrial Communication Systems}. In Proc. of ICST on Personal Satellite Services, Feb 2011, Malaga.

\bibitem{PIMRC}
S. Kandeepan, K. Gomez, T. Rasheed and L. Reynaud, \emph{energy-efficient Cooperative Strategies in Hybrid {Aerial-Terrestrial} Networks for Emergencies}. In Proc. of IEEE PIMRC, Toronto, Canada, pp. 173-178, 2011.

\bibitem{E2NETS12}
L. Reynaud, K. Sithamparanathan, K. Gomez and T. Rasheed, \emph{Adaptive energy-efficient Communications for Hybrid Aerial-Terrestrial Systems}. in Proc. IEEE ICC 2012 Workshop (E2Nets), Budapest, Hungary, 2013.

\bibitem{Haykin}
S.~Haykin, \emph{Cognitive Radio: Brain-Empowered Wireless Communication}. IEEE Journal on Selected Areas in Communications, vol.~23, no.~2, pp.201-220, Feb. 2005.

\bibitem{C2POWER_website}
C2POWER Project; European Commission funded, EU-FP7 STREP Project (ICT-248577)). Available at \url{http://www.ict-c2power.eu}.


\bibitem{relay2}
R. Tannious and A. Nosratinia, \emph{Spectrally-efficient relay selection with limited feedback}. IEEE Journal on Selected Areas in Communications, 2008, 26(8): 1419-1428

\bibitem{relay5}
A. Bletsas, A. Khisti and D. R. Reed, \emph{A simple cooperative diversity method based on network path selection}. IEEE Journal on Selected Areas in Communications, 2006, 24(3): 659-672.

\bibitem{relay6}
M. Elhawary and Z.J. Haas, \emph{Energy-Efficient Protocol for Cooperative Networks}. IEEE/ACM Transactions on Networking, vol.19, no.2, pp.561-574, April 2011.

\bibitem{new_ref1}
Zhong Zhou, Shengli Zhou, Jun-Hong Cui and Shuguang Cui, Energy-Efficient Cooperative Communication Based on Power Control and Selective Single-Relay in Wireless Sensor Networks, IEEE Transactions on Wireless Communications, Vol. 7, No. 8, August 2008.

\bibitem{new_ref2}
Beibei Wang, Zhu Han and K.J. Ray Liu, Distributed Relay Selection and Power Control for Multiuser Cooperative Communication Networks Using Stackelberg Game, IEEE Transactions on Mobile Computing, Vol. 8, No. 7, July 2009

\bibitem{relay4}
R. Madan, N. B. Mehta and Molisch A F, \emph{Energy-efficient cooperative relaying over fading channels with simple relay selection}. IEEE Transaction on Wireless Communications, 2008, 7(8): 3013-3025.

\bibitem{relay1}
Y. Wei, L. Li-hua, and S. Wan-lu, \emph{Energy-efficient relay selection and optimal relay location in cooperative cellular networks with asymmetric traffic}. The Journal
of China Universities of Posts and Telecommunications Published by Elsevier B.V., vol. 17, pp. 80-88, 2010.

\bibitem{PControl1}
R. Annavajjala, P. C. Cosman, and L. B. Milstein, Statistical Channel Knowledge-Based Optimum Power Allocation for Relaying Protocols in the High SNR Regime, IEEE Journal on Selected Areas In Communications, Vol. 25, No. 2, February 2007, pp 292-305.

\bibitem{PControl2}
Y. Zhu, Y. Xin, and Pooi-Yuen Kam, Optimal Transmission Strategies for Rayleigh Fading Relay Channels, IEEE Transactions on Wireless Communications, Vol. 7, No. 2, February 2008, pp 618-628.

\bibitem{PControl3}
T. T. Pham, H. H. Nguyen and H. D. Tuan, Power Allocation in Orthogonal Wireless Relay Networks With Partial Channel State Information, IEEE Transactions on Signal Processing, Vol. 58, No. 2, February 2010, pp. 869-878.

\bibitem{PControl4}
G. Farhadi and N. C. Beaulieu, Power-Optimized Amplify-and-Forward Multi-Hop Relaying Systems, IEEE Transactions on Wireless Communications, Vol. 8, No. 9, September 2009, pp 4634-4643.

\bibitem{PControl5}
A. Khabbazibasmenj and S. A. Vorobyov, Power Allocation Based on SEP Minimization in Two-Hop Decode-and-Forward Relay Networks, IEEE Transactions on Signal Processing, Vol. 59, No. 8, August 2011, pp 3954-3963.

\bibitem{PControl6}
I. Hammerstrom and A. Wittneben, On the Optimal Power Allocation for Nonregenerative OFDM Relay Links, IEEE ICC 2006, Istanbul.

\bibitem{PControl7}
B. Wang, Z. Han and K. J. Ray Liu, Distributed Relay Selection and Power Control for Multiuser Cooperative Communication Networks Using Buyer/Seller Game, IEEE INFOCOM 2007, Alaska.

\bibitem{PTrading}
S. Kandeepan, S. K. Jayaweera, and R. Fedrizzi, ‘Power-Trading in Wireless Communications: A cooperative Networking Business Model’,  IEEE Transactions of Wireless Communications, vol.11, iss.5, 2012.

\bibitem{PControl8}
B. Wang, Z. Han and K.J. Ray Liu, Distributed Relay Selection and Power Control for Multiuser Cooperative Communication Networks Using Stackelberg Game, IEEE Transactions on Mobile Computing, Vol. 8, No. 7, July 2009, pp 975-990.

\bibitem{PControl9}
Feng Li, and Li Wang, A Dynamic Game Algorithm for Power Allocation in Cognitive Relay Transmission, IEEE Transactions on Communications, Vol. 61, No. 11, November 2013, pp. 4440-4447.

\bibitem{gps}
J.H. Blakey, \emph{Navigating towards the future: transitioning from terrestrial radio navigation to satellite navigation and airborne surveillance}. IEEE Aerospace and Electronic Systems Magazine, vol.21, no.5, pp.17,21, May 2006.

\bibitem{gps1}
P. Pech, H. Puming, M. Bousquet, \emph{Low bit rate satellite link for emergency communications}. IEEE International Workshop on Satellite and Space Communications, October 2008, Toulouse.

\bibitem{HAP_Book2}
D. Grace and M. Mohorcic, \emph{Broadband Communications via High-Altitude Platforms}. Ed. John Wiley and Sons, UK, 2010.

\bibitem{Project1}
FP6 CAPANINA Project. Available at \url{http://www.capanina.org}.

\bibitem{Project2}
HAPCOS COST 297. Available at \url{www.hapcos.org}.

\bibitem{PACE_ALOI}
P. Pace and G. Aloi, \emph{Disaster monitoring and mitigation using aerospace technologies and integrated telecommunication networks}, IEEE Aerospace and Electronic System Magazine, vol. 23, Apr. 2008, pp. 3-9.

\bibitem{HAP_energy}
L. Zhu, X. Yan and Y. Zhu, \emph{High altitude platform-based two-hop relaying emergency communications schemes}. In Proc. of Wireless communications, networking and mobile computing (WiCOM'09), Beijing, China, 2009.

\bibitem{HAP_antenna}
S. Denton, P. Zavidniak and W. Hodgkiss, \emph{High altitude airborne cellular base stations antenna analysis}.  IEEE Aerospace and Electronic Systems Magazine, vol.25, no.4, pp.30,36, April 2010.

\bibitem{Hybrid}
S. Aranti, M. De Sanctis, S.C. Spinella, M. Monti, E. Cianca, A. Molinaro, A. Iera and M. Ruggieri, \emph{Hybrid System HAP-WiFi for Incident Area Networks}. In Proc. of ICST Personal Satellite Services, pp. 436-450, February 2010.

\bibitem{Indonesia_ATN}
A. Qiantori, B. Sutiono, H. Hariyanto, H. Suwa and T. Ohta, \emph{An Emergency Medical Communications System by Low Altitude Platform at the Early Stages of a Natural Disaster in Indonesia}, \emph{International Journal of Medical Systems}, vol. 34, March 2010, pp. 1-12.

\bibitem{uav1}
D. Erdos, A. Erdos and S.E. Watkins, \emph{An experimental UAV system for search and rescue challenge}. IEEE Aerospace and Electronic Systems Magazine, vol.28, no.5, pp.32,37, May 2013.

\bibitem{uav2}
E. Pastor, J. Lopez and P. Royo, \emph{UAV Payload and Mission Control Hardware/Software Architecture}. IEEE Aerospace and Electronic Systems Magazine, vol.22, no.6, pp.3,8, June 2007.

\bibitem{gossip}
Z. Haas, J. Halpern and L. Li, \emph{Gossip-based ad hoc routing}. In Proc. of IEEE INFOCOM, 2002, New York.

\bibitem{gossip1}
R. Beraldi, \emph{"The polarized gossip protocol for path discovery in MANETs"}, Ad Hoc Networks, vol. 6, pp. 79-91, 2008.

\bibitem{gossip2}
B. Krishnamachari, S.B. Wicker and R.B., M. Pearlman, \emph{Critical Density Thresholds in Distributed Wireless Networks}. In Proc. of Communications, Informations and Network Security, December, 2002.

\bibitem{gossip3}
X.-Y. Li, K. Moaveninejad and O. Frieder, \emph{Regional Gossip Routing for Wireless Ad Hoc Networks, Mobile Networks and Applications (MONET)}, volume 10, February, 2005, pp. 61-77.
,
\bibitem{gossip4}
J. Luo, P. Th. Eugster and J.-P. Hubaux, \emph{Route Driven Gossip: Probabilistic Reliable Multicast in Ad Hoc Networks}. In Proc. of INFOCOM, 2003, San Frasisco.

\bibitem{comp1}
A. Miyoshi, C. Lefurgy, E. Hensbergen, R. Rajamony, and R. Rajkumar, \emph{Critical power slope: understanding the runtime effects of frequency scaling}. In Proc. of Supercomputing. ACM, New York, NY-USA, 2002.

\bibitem{comp2}
A. Carroll and G. Heiser, \emph{An analysis of power consumption in a smartphone}. In Proc. of USENIXATC. USENIX Association, Berkeley, CA-USA, 2010.

\bibitem{comp3}
A. Pathak, Y. Charlie Hu, M. Zhang, P. Bahl, and Y. Wang, \emph{Fine-grained power modeling for smartphones using system call tracing}. In Proc. of EuroSys. ACM, New York, NY-USA, 2011.

\bibitem{GREENCOM1}
K. Gomez, D. Boru, R. Riggio, T. Rasheed, D. Miorandi, and F. Granelli, \emph{Measurement-based modelling of power consumption at wireless access network gateways} Computer Networks, vol.56, pp.2506-2521, July 2012.


\bibitem{JSAC_channelmodel}
F. Dovis, R. Fantini, M. Mondin, and P. Savi, \emph{Small-Scale Fading for High-Altitude Platform (HAP) Propagation Channels}. IEEE Journal on Selected Areas in Comm., Vol20, No3, April 2002, pp 641-647.

\bibitem{Rapp}
T. S. Rappaport, \emph{Wireless Communications: Principles and Practice}. Englewood Cliffs, NJ: Prentice-Hall, 1996.

\bibitem{MastersThesis}
L. C. Liechty, Path Loss Measurements and Model Analysis of a 2.4 Ghz Wireless Network in an Outdoor Environment, Master Thesis, Georgia Institute of Technology, 2007.

\bibitem{alpha}
E. Frewand and T. Brown, \emph{Airborne communication networks for small unmanned aircraft systems}. In Proc. of the IEEE, vol. 96, no. 12, pp. 2008-2027, December 2008.

\bibitem{kandee1}
A. Hourani, S. Kandeepan, ‘Cognitive Relay Nodes for Airborne LTE Emergency Network’ IEEE ICSPCS, Dec 2013, Gold Coast.

\bibitem{kandee2}
K. Gomez, T. Rasheed, L. Reynaud and S. Kandeepan, On the Performance of Aerial LTE Base-Stations for Public Safety and Emergency Recovery, IEEE Globecom, WiUAV-WS, December, 2013, Atlanta.

\bibitem{kandee3}
S. Chandrasekharan, S. Kandeepan, et.al, Clustering Approach for Aerial Base-Station Access with Terrestrial Cooperation, IEEE Globecom, WiUAV-WS, Dec 2013, Atlanta.

\bibitem{kandee4}
A. Hourani, S. Kandeepan, ‘Temporary Cognitive Femtocell Network For Public Safety LTE, IEEE CAMAD, Sep 2013, Berlin.

\bibitem{kandee5}
S. Arunthavanathan, S. Kandeepan, R. Evans, Spectrum Sensing and Detection of Incumbent-UEs in Secondary-LTE based Aerial-Terrestrial Networks for Disaster Recovery. IEEE CAMAD, Sep 2013, Berlin.

\bibitem{kandee6}
A. Alvaro, et. al, ‘Airborne Base Stations for Emergency and Temporary Events’, Int conf on. Personal Satellite Services (PSATS), Springer, Feb 2013, Touluse

\bibitem{kandee7}
S. Kandeepan et. al, “Spectrum Sensing for Cognitive Radios with Primary User Transmission Statistics: with Linear Frequency Sweeping”, JWCN EURASIP, Special issue on Dynamic Spectrum Access: From the Concept to the Implementation, May 2010

\bibitem{kandee8}
A. Mariani, S. Kandeepan, A. Giorgetti, M. Chiani, 'Cooperative Weighted Centroid Localization for Cognitive Radio Networks', To appear in Proceedings of International Symposium on Communications and Information Technologies (ISCIT), 2-5 October, 2012, Gold Coast.

\bibitem{kandee10}
S. Kandeepan et. al, “Distribute ring-around Spectrum Sensing for Cognitive Radio Networks”, IEEE ICC 2011, Kyoto

\bibitem{kandee11}
S. Kandeepan, L. Nardis, M. G. Benedetto, G. Corazza, Alessandro, “Cognitive Satellite Terrestrial Radios”, IEEE Globecom, Dec 2010, Florida,

\bibitem{kandee12}
S. Kandeepan et. al, “Time-Divisional Cooperative Periodic Spectrum Sensing for Cognitive Radio Networks”, IEEE ICC, 23-27 May 2010, Cape Town

\bibitem{kandee13}
S. Kandeepan et. al, “Periodic Sensing in Cognitive Radios for Detecting UMTS/HSDPA Based on Experimental Spectral Occupancy Statistics”, IEEE WCNC, April 2010

\bibitem{kandee14}
Kandeepan. S, et al, “Bayesian Tracking in Cooperative Localization for Cognitive Radio Networks”, IEEE VTC , 26-29 April, Barcelona 2009

\bibitem{kandee15}
S. Kandeepan et. al, “Experimentally Detecting IEEE 802.11n Wi-Fi Based on Cyclostationarity Features for Ultra-Wide Band Cognitive Radios”, IEEE PIMRC 2009, Tokyo

\bibitem{kandee16}
S. Kandeepan et. al, “Periodic Spectrum Sensing Performance and Requirements for Detecting Legacy Users with Temporal and Noise Statistics in Cognitive Radios”, IEEE BWA-WS Globecom, Dec 2009, Honolulu

\bibitem{kandee17}
Rahim, T. Aysal, Kandeepan. S, Dzmitry. K, Radoslaw. P, “Cooperative Shared Spectrum Sensing for Dynamic Cognitive Radio Network”, International Conference on Communications (ICC), 14-18 June , Dresden 2009

\bibitem{kandee18}
S. Kandeepan, et. al, “Preliminary Experimental Results on the Spectrum Sensing Performances for UWB-Cognitive Radios for Detecting IEEE 802.11n Wi-Fi Systems”, IEEE ISWCS, Sienna, Sep 2009

\bibitem{kandee19}
S. Kandeepan, et. al, “Time-Divisional and Time-Frequency Divisional Cooperative Spectrum Sensing”, IEEE Crowncom 09, Hannover, June 2009

\bibitem{kandee20}
T. Aysal, Kandeepan. S, Radoslaw. P, “Cooperative Spectrum Sensing with Noisy Hard Decision Transmissions”, International Conference on Communications (ICC), 14-18 June , Dresden 2009

\bibitem{OMNET}
OMNET++ Network Simulator. Available at: \url{http://www.omnetpp.org}.


\end{thebibliography}
\end{document}